\pdfoutput=1
\documentclass{JINST}
\usepackage{graphicx}
\usepackage{amssymb}

\title{New approach to 3D electrostatic calculations for micro-pattern detectors}

\author{Predrag Lazi\'{c}$^a$\thanks{Corresponding author.}~, Denis Dujmi\'{c}$^{b,c}$, Joseph A. Formaggio$^{b,c}$, Hrvoje Abraham$^d$ and Hrvoje \v{S}tefan\v{c}i\'{c}$^e$\thanks{The work of Hrvoje \v{S}tefan\v{c}i\'{c} on this paper was performed outside his working hours at Rudjer Bo\v{s}kovi\'{c} Institute and outside facilities of Rudjer Bo\v{s}kovi\'{c} Institute.} \\
\llap{$^a$}Department of Materials Science and Engineering,\\
  Massachusetts Institute of Technology, Cambridge, Massachusetts 02139, USA\\
\llap{$^b$}Laboratory for Nuclear Science,\\
  Massachusetts Institute of Technology, Cambridge, Massachusetts 02139, USA\\
\llap{$^c$}Department of Physics,\\
  Massachusetts Institute of Technology, Cambridge, Massachusetts 02139, USA\\
\llap{$^d$}Artes Calculi,\\
  Fra Grge Marti\'{c}a 24, Zagreb, Croatia \\ 
\llap{$^e$}Theoretical Physics Division,\\
  Rudjer Bo\v{s}kovi\'{c} Institute, POB 180, HR-10002 Zagreb, Croatia \\ 
  E-mail: \email{plazic@mit.edu}}

\abstract{We demonstrate practically approximation-free electrostatic calculations of micromesh detectors that can be extended to any other type of micropattern detectors. Using newly developed Boundary Element Method called Robin Hood Method we can easily handle objects with huge number of boundary elements (hundreds of thousands) without any compromise in numerical accuracy. In this paper we show how such calculations can be applied to Micromegas detectors by comparing electron transparencies and gains for four different types of meshes. We demonstrate inclusion of dielectric material by calculating the electric field around different types of dielectric spacers.}


\begin{document}

\section{Introduction}
\begin{figure}[htb]
\begin{center}
\includegraphics[width=1.0\linewidth,clip=true]{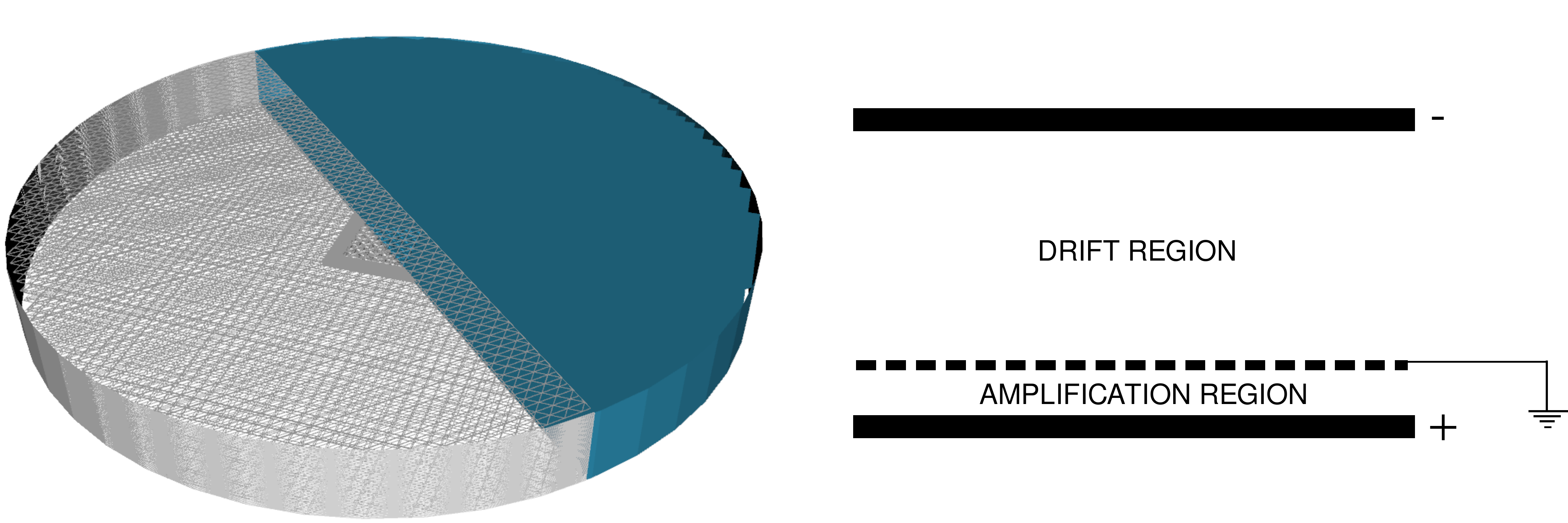}
\end{center}
\caption{The whole detector as calculated - left. Schematic picture of the detector - right. A charged particle ionizes gas in the drift region leaving a trail of primary ionization electrons. In order to read them out, the electrons drift in a weak field through a micromesh, and enter the high-field amplification region where they undergo  avalanche multiplication.}
\label{whole_detector}
\end{figure}

It has been more than two decades since a new type of gaseous detector was proposed, characterized by a small distance between electrodes in the charge amplification region (tens to hundreds of $\mu$m) \cite{megas_1}. Nowadays, these so called micro-pattern detectors (in most widely used versions as Micromegas \cite{charpak} and GEMs \cite{megas_2}) are established as good replacements for the multiwire proportional chamber due to finer granularity, ability to operate at high ionization rates and simple manufacturing. Their use ranges from trackers in accelerator-based experiments, large time-projection chambers in rare-event detectors to radiation detectors \cite{megas_3}.
Despite the widespread acceptance of micro-pattern detectors, a precise calculation of the electric field inside the three-dimensional structure of electrodes is still a challenge. In this paper we choose a Micromegas geometry as a proxy for all micro-pattern detectors, and perform three dimensional electric field calculation. We compare different mesh geometries and expected performances.\\
A micromesh in Micromegas detectors \cite{charpak} separates the low-field conversion volume from the high-field amplification region (see figure \ref{whole_detector}). The electric field in this transitional volume depends on the geometric shape of the micromesh and affects the performance of the detector. Calculation of the electric field is difficult due to three-dimensional nature of the problem and high variation of the field in the small volume around the mesh. This requires a fine spatial fragmentation and leads to computational complexity that presents a problem for popular algorithms. A standard remedy is to simplify the detector geometry, e.g. by ignoring the third dimension \cite{garfield} or approximating the mesh geometry \cite{simplify_mesh}.
We present a calculation without such compromises that uses a novel BEM algorithm \cite{RH1,RH2}. We calculate electronic transparency and gain for four mesh types and vary the pitch of the mesh and ratios of the drift and amplification field. Also we explore the influence of the dielectric spacer on the electric field in the device \cite{denis_direction_sensitive}.

\section{Theory and methods}
Principles of operation of Micromegas detectors are tightly connected to electrostatic problem with a set of electrodes at different applied voltages. Given the boundary conditions one searches for the electric field inside a detector in order to simulate a trajectory of a charged particle (electron). The ionization, electron attachment and diffusion are taken into account using tabulated values for CF$_4$ gas at 760 Torr of pressure \cite{Magboltz}.

\subsection{Electrostatics and Boundary Element Methods}
Since the most important quantity in studying the operation of Micromegas detector is the electric field our method of choice for electrostatic part of the problem is the Boundary Element Method (BEM) \cite{RH1,RH2,indijci_1_electrostatics_3d}. BEM in electrostatics consists of finding the charge density distribution on a set of given boundary elements (usually triangles) that satisfy the given boundary conditions - i.e. potential values at electrodes. Once the charge densities are found owing to the Green's function for the Coulomb interaction we can reconstruct potential and electric field accurately anywhere in space. Most of the existing electrostatic codes are based on a finite differences or finite element methods \cite{maxwell_3d,comsol} which find a potential values on a grid of points in volume respecting the given boundary conditions on electrodes. Besides the fact that such solutions are not solutions to the Maxwell equations \cite{veenhof_maxwell} there is another problem that the grid of points has to be terminated artificially somewhere in space \cite{indijci_2_truncate_FEM}, introducing an arbitrary parameter into calculations. Moreover, an impractically fine density of grid points is required in order to achieve accurate values of the electric field around corners and edges of electrodes. Additionally inaccuracies in the electric field values add up along the simulated trajectory of the particle.
In classical BEM implementation for electrostatics, the biggest problem is the size of the Coulomb interaction matrix. This matrix is not sparse and explodes in size as $N^2$ with the number of boundary elements. This limits the calculation to a mere few thousands of boundary elements \cite{indijci_1_electrostatics_3d, indijci_2_truncate_FEM, indijci_3_realistic_3d}, which restricts the complexity of problems and hinders the wider usage of BEM in electrostatics. In this work, we use a novel algorithm named Robin Hood Solver (RH) \cite{solver} which has elegantly solved this challenge of matrix size without a compromise in accuracy. This enables us to treat systems of several hundreds of thousands of triangles on a single CPU with small amount of memory with ease (within several hours). 
The RH code itself does not include a simulation of electron propagation so one option is to use the \textit{Garfield} code \cite{garfield} and supply the electric field map calculated by the RH.
However, we have decided to make our own implementation of the codes to calculate transparency and gain in order to have greater control over the output of the code.
\section{Computations and results}

\begin{figure*}[htb]
\begin{center}
{
\includegraphics[width=0.22\linewidth,clip=true]{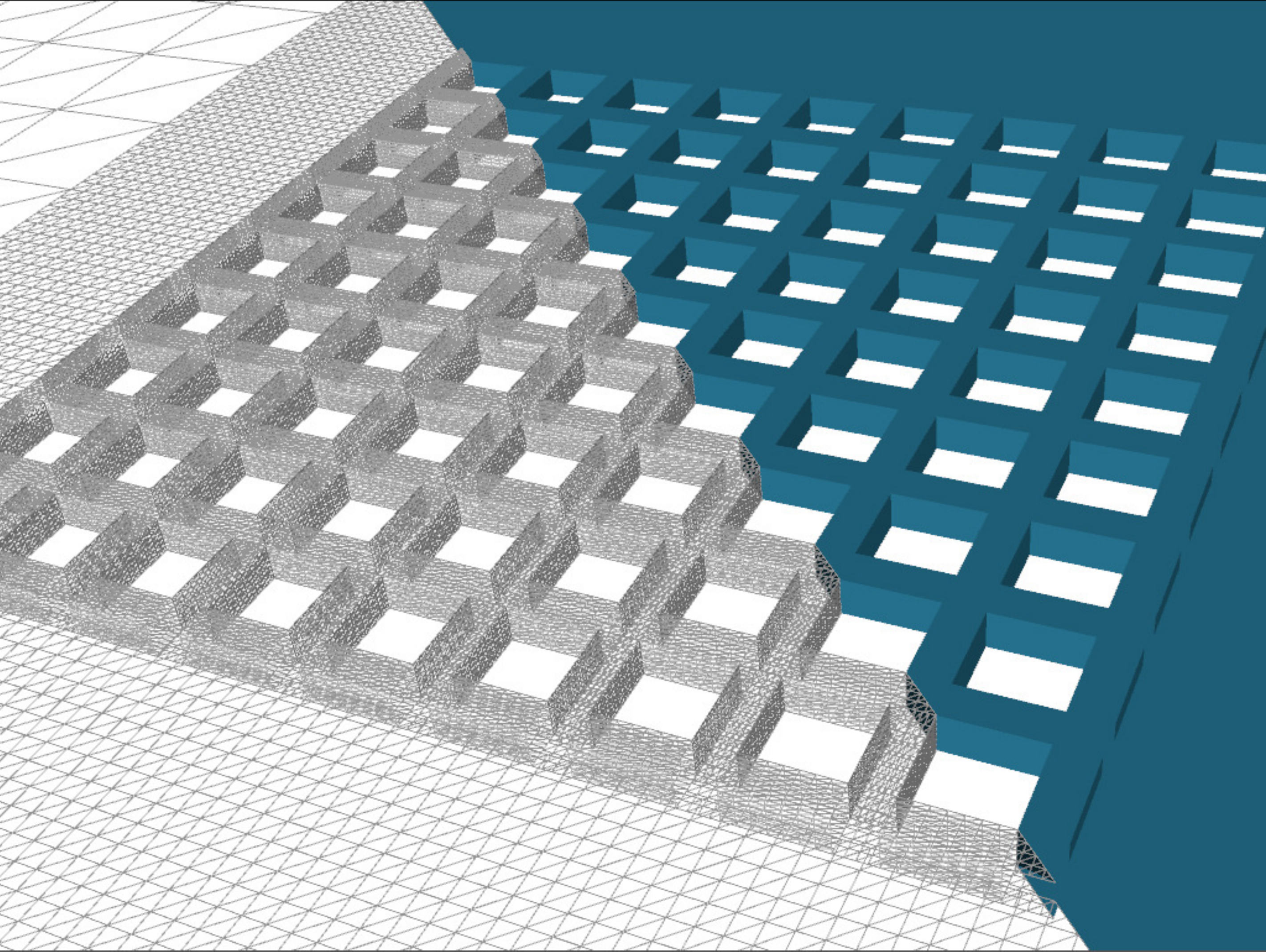}
\includegraphics[width=0.22\linewidth,clip=true]{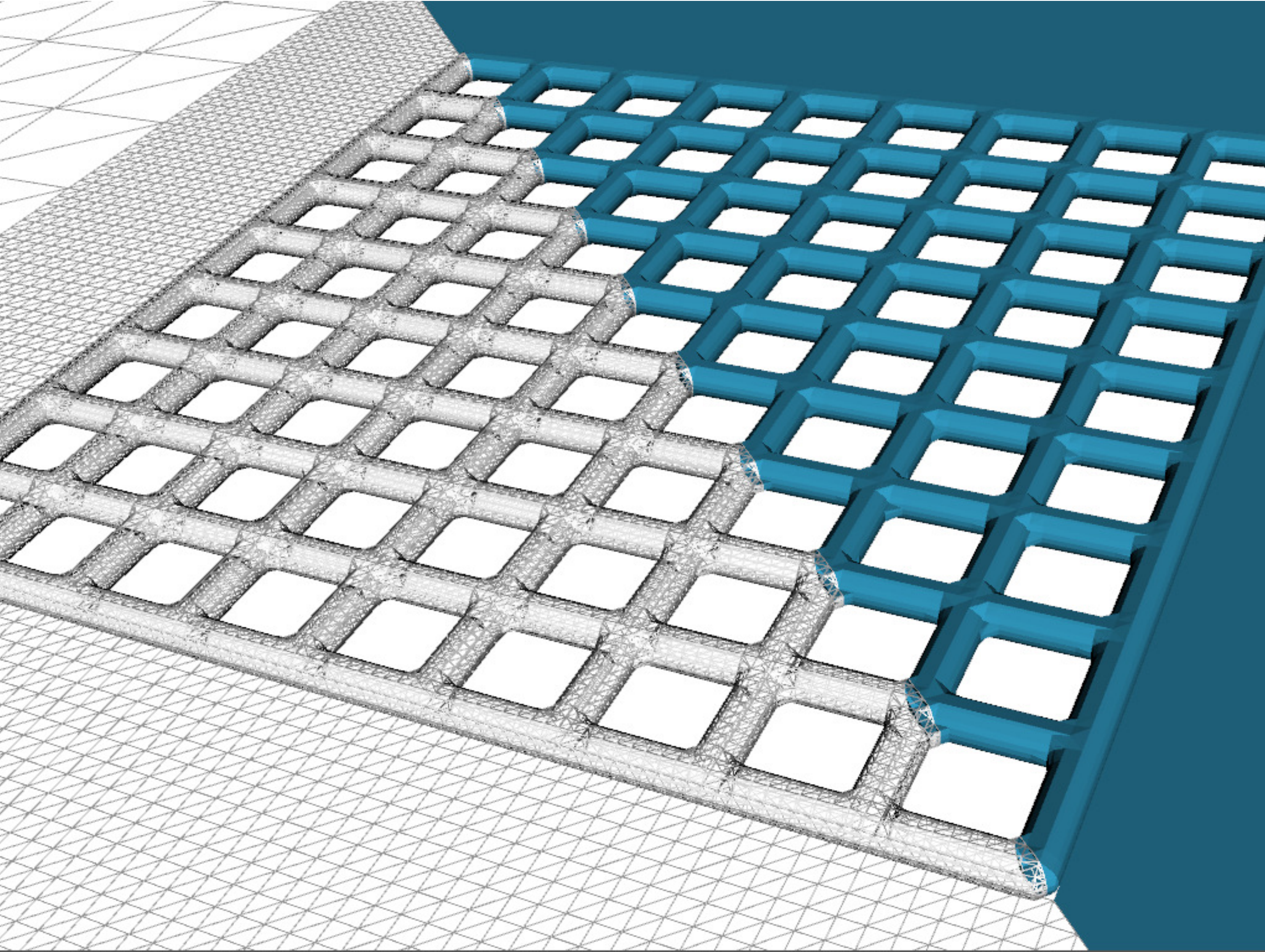}
\includegraphics[width=0.22\linewidth,clip=true]{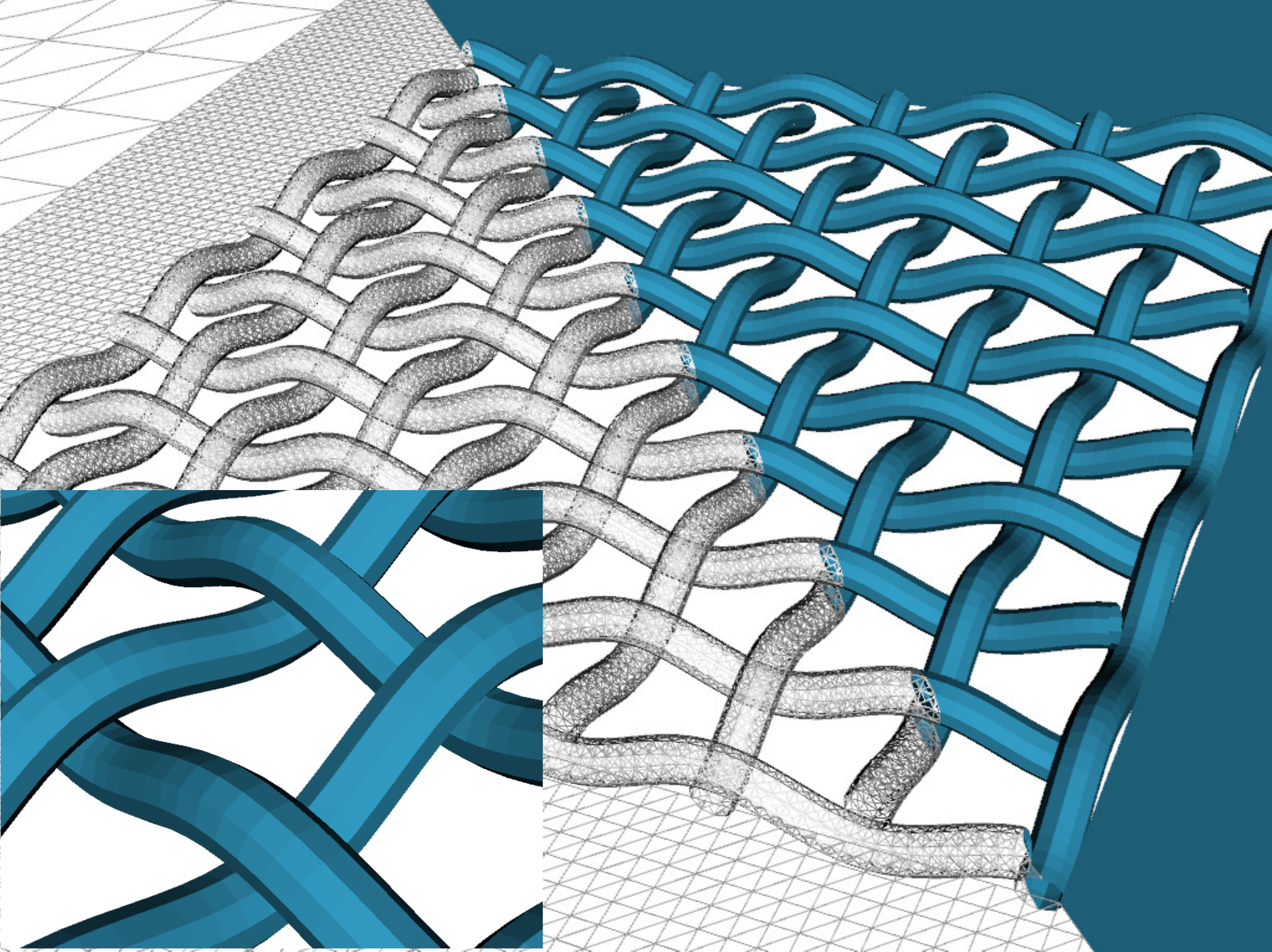}
\includegraphics[width=0.22\linewidth,clip=true]{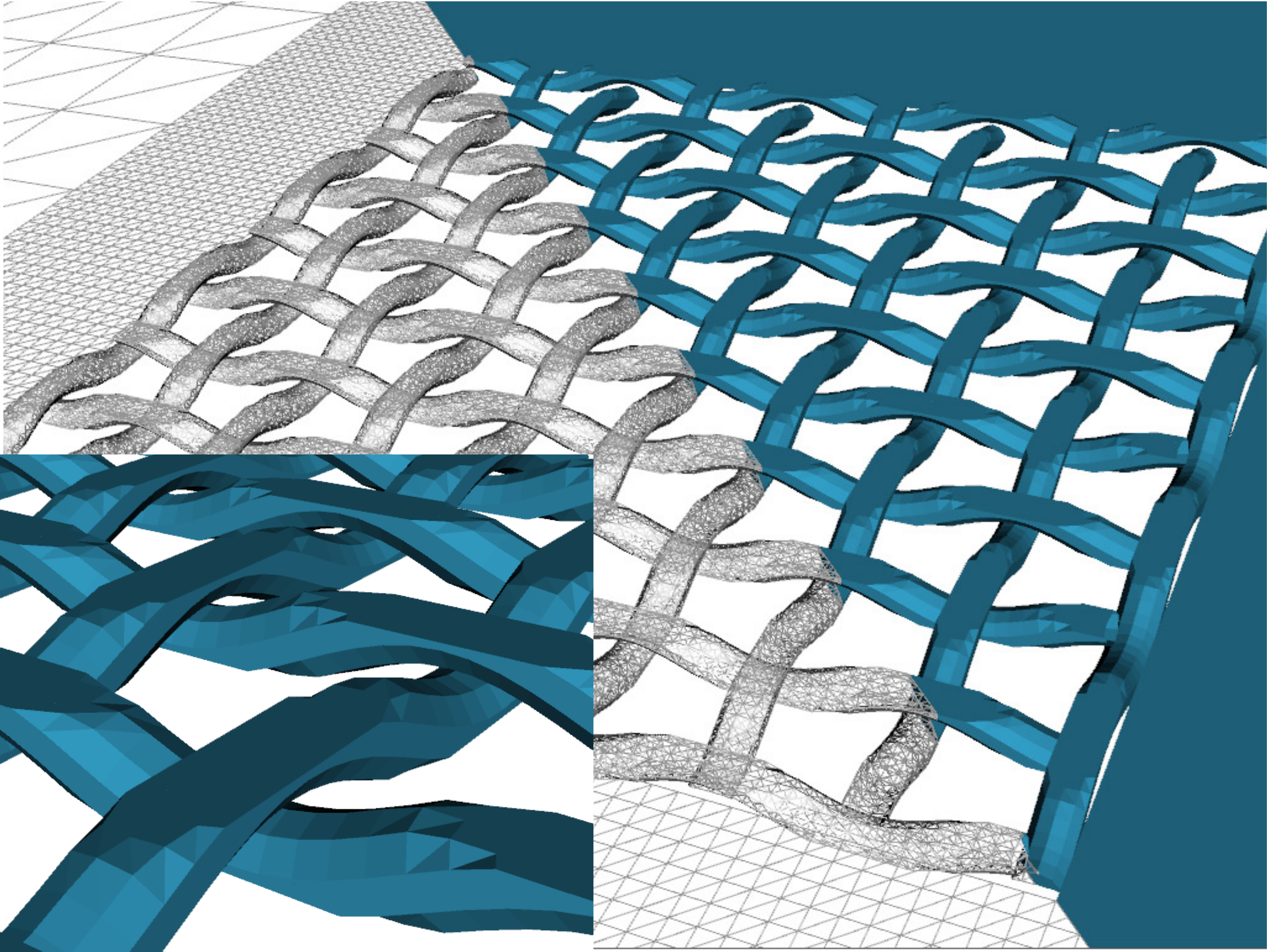}
}
\end{center}
\caption{Different types of micro-mesh, from left to right: rectangular, cylindrical, woven and calendered.}
\label{meshes}
\end{figure*}

The four types of micro-meshes that we use in our calculations are shown in figure \ref{meshes}. We named them as follows: rectangular, cylindrical, woven and calendered micro-meshes.

We have chosen 8 different amplification-to-drift field ratios, 8 different mesh pitches while keeping in all cases the diameter of the wire constant d=28.2 microns (in the case of a rectangular mesh this is the width of the wire). In our model of a detector micro-mesh is always positioned so that its center is at 50 microns above the bottom electrode and all voltages are fixed except the voltage of the bottom electrode (anode). By varying the voltage we achieve different drift-to-amplification field ratios. The total size of the mesh area is 2.5mm $\times$ 2.5mm that is much larger than the scale of the signal spread due to e.g. electron diffusion. 
The number of triangles for a different type of meshes and different pitches varies a bit but is never smaller than 300,000. The equipotentiality of the electrodes is satisfied always to 1 part in 10$^4$ in our RH solutions. \\
In the first subsection we calculate the transparency and the gain, while in the second subsection we look at the influence of the dielectric spacer placed between the mesh and the anode. 
We use three different types of spacers - a horizontal filled cylinder, a horizontal hollow cylinder (capillary), and a vertical hollow cylinder - in order to study the spacer influence on the electric field inside the detector.

\subsection{Transparency and gain}
\begin{figure}[htb]
\begin{center}
{
\includegraphics[width=0.45\linewidth,clip=true]{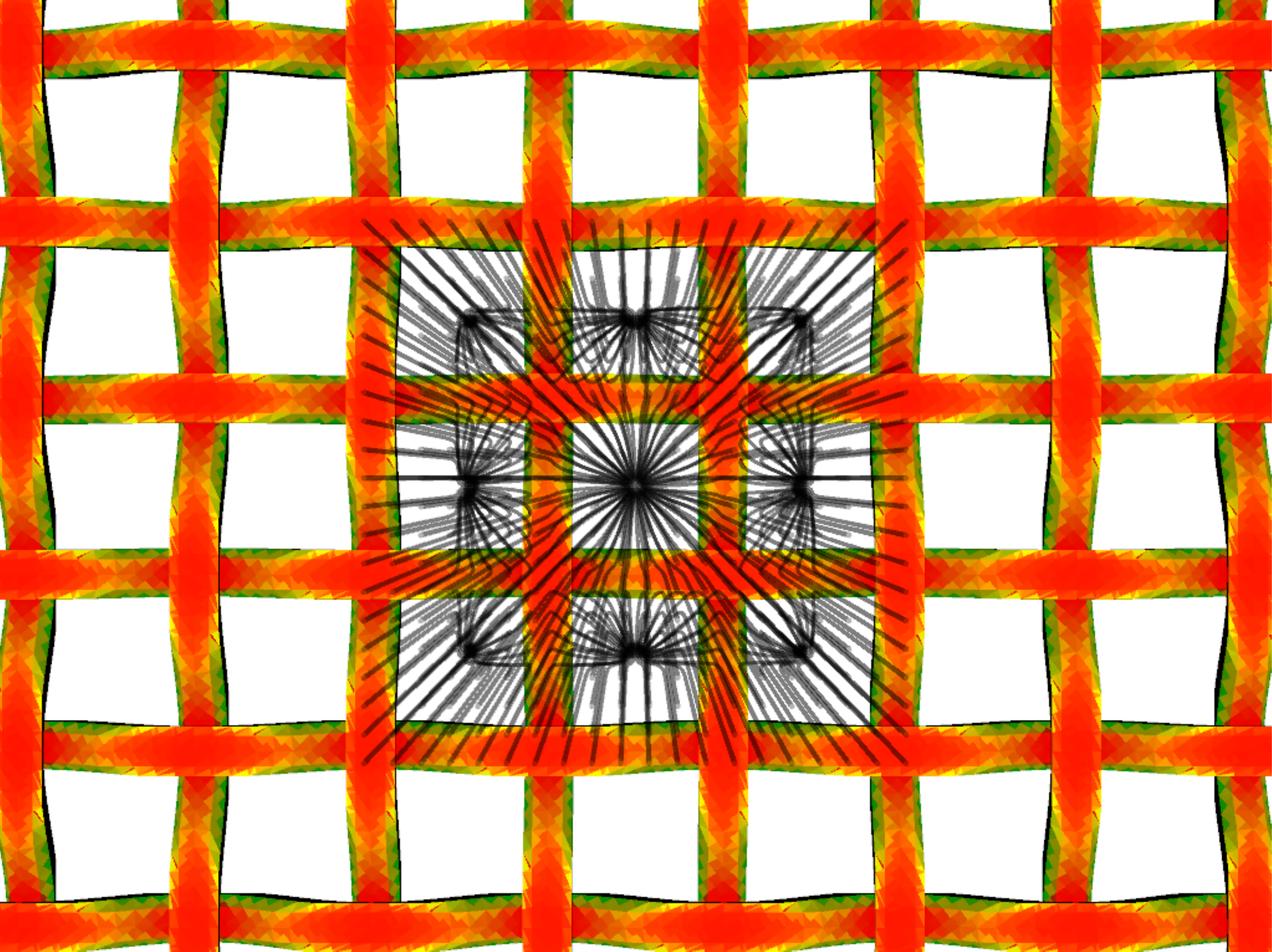}
\includegraphics[width=0.45\linewidth,clip=true]{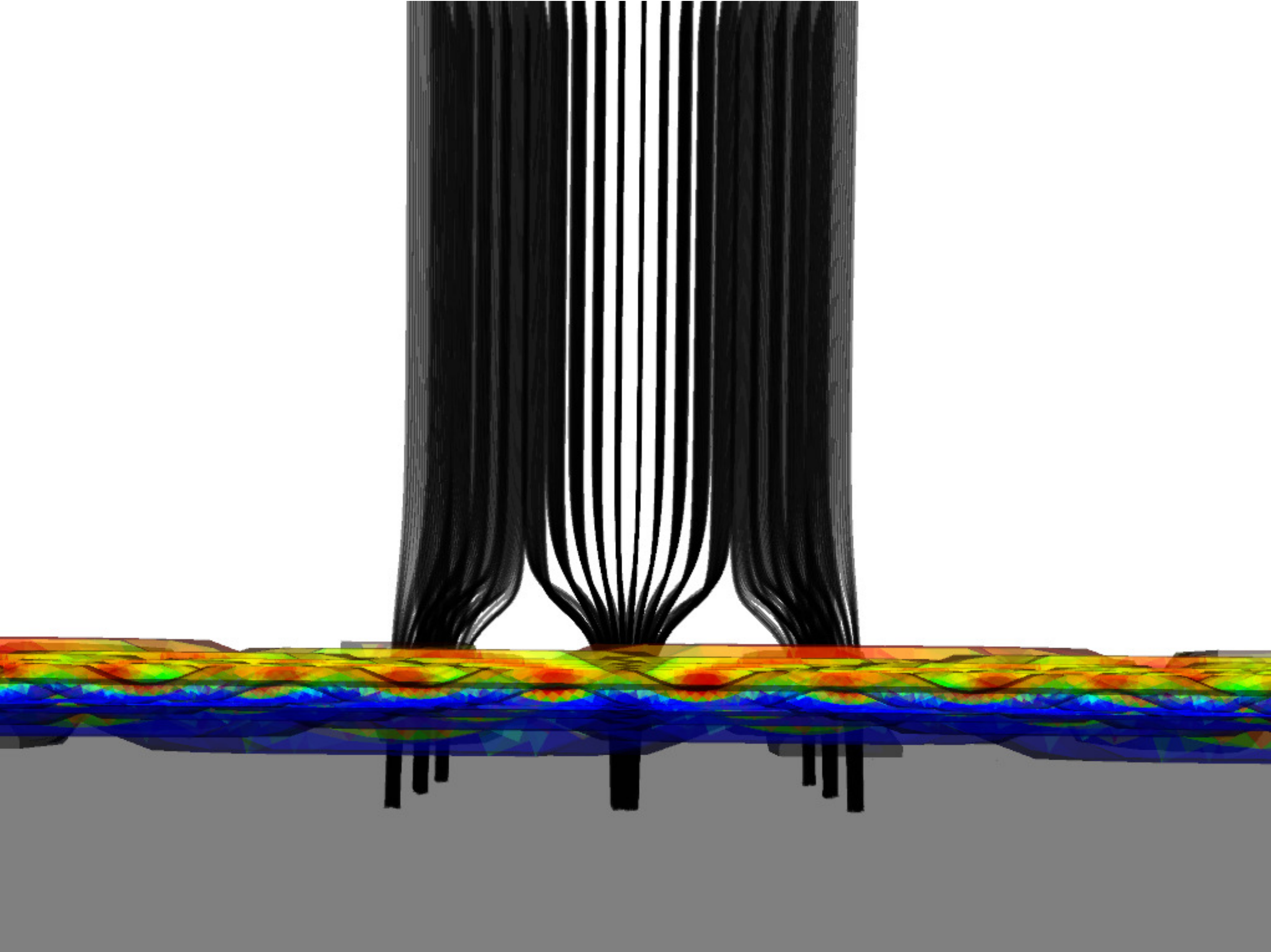}
\includegraphics[width=0.45\linewidth,clip=true]{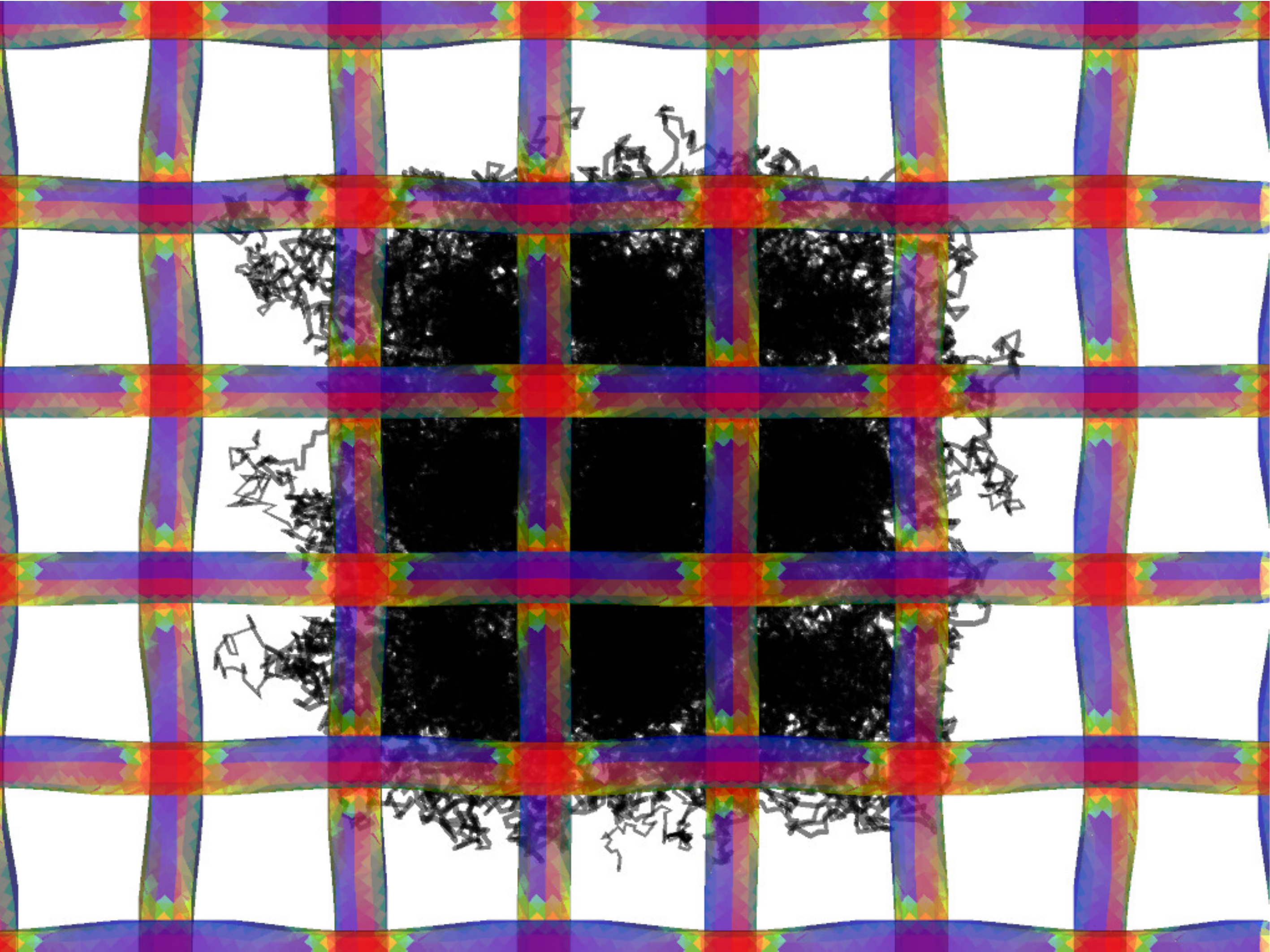}
\includegraphics[width=0.45\linewidth,clip=true]{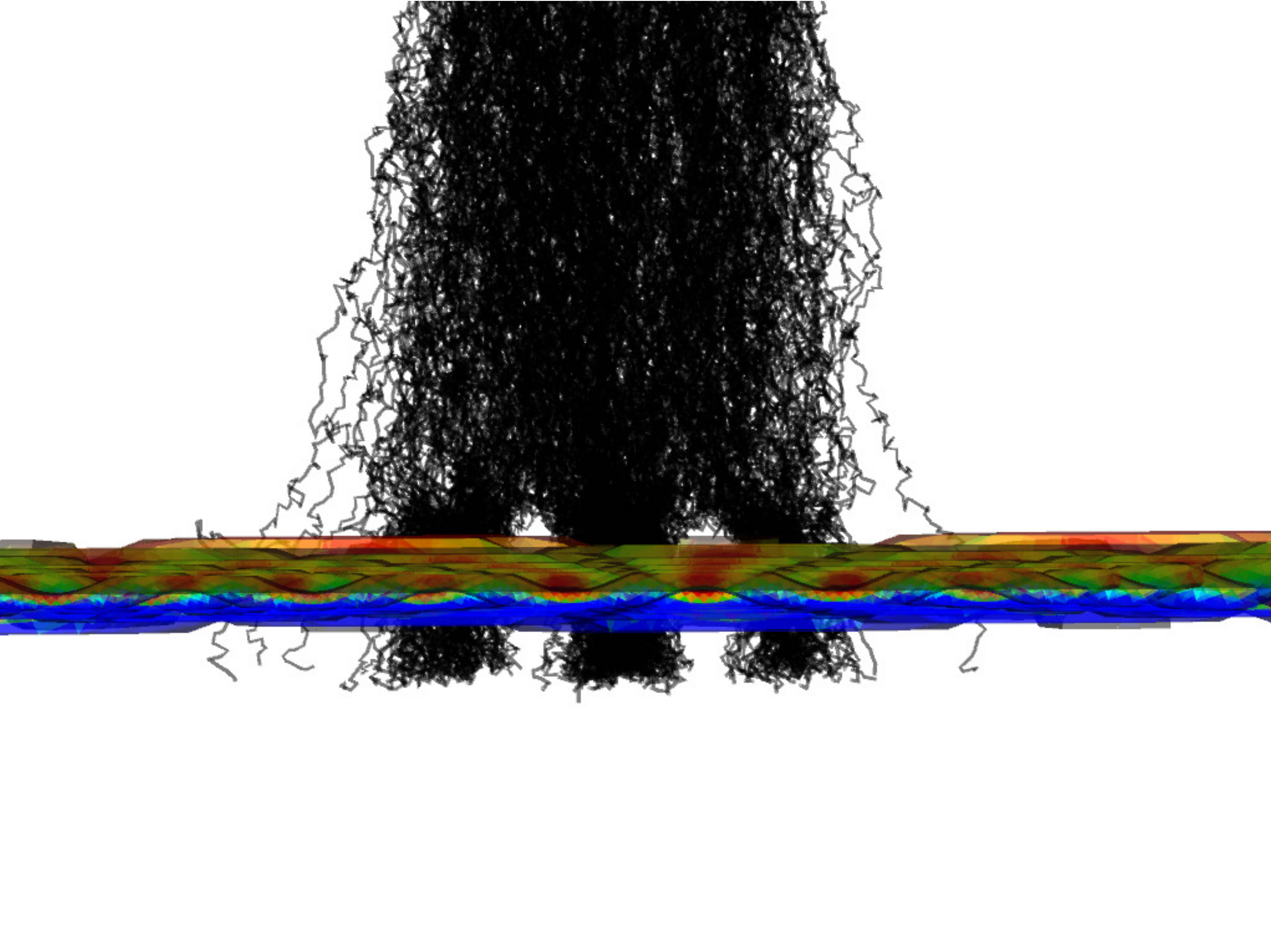}
}
\end{center}
\caption{Plots of electron trajectories that follow electric field lines are shown in the two images on the top. Plots of microscopic electron tracking that take into account the electron diffusion are shown in the bottom.}
\label{lines_of_force}
\end{figure}

The transparency and gain are calculated using trajectories of primary-ionization electrons that are randomly created in the drift region. 
The transparency is defined as the fraction of primary-ionization electrons that reach the amplification region.  
In order to calculate electron trajectories we use two methods. In the electric-field tracking (EFT) method, we ignore the electron diffusion in gas, and place electron trajectories along the lines of the electric field flux \cite{simplify_mesh}. 
The electric field is calculated using charges on the electrodes, which is more accurate than using interpolation from a set of electric field points. In the electron micro-tracking (MT) method \cite{simplify_mesh}, we include the effects of the diffusion using tabulated values for diffusion coefficients.
In both tracking approaches, we calculate the gain as the integral of ionization and attachment coefficients along the electron trajectory. 
For each calculation we sample 400 trajectories. Plots of trajectories for EFT and MT calculations are shown in figure \ref{lines_of_force}.\\

In all cases, a more realistic MT calculation yields a smaller transparency and a smaller gain than the EFT. Among the four studied types of meshes cylindrical and calendered are the most transparent ones - actually those two meshes have, somewhat to our surprise, very similar transparencies across the whole range of calculated parameters. Rectangular mesh turns out to be the worst of all 4 cases yielding the lowest electrical transparency. Woven mesh ends up somewhere in between the case of a bad mesh (rectangular) and good meshes (cylindrical and calendered) regarding its transparency.

\begin{figure*}[htb]
\begin{center}
{
\includegraphics[width=0.23\linewidth,clip=true]{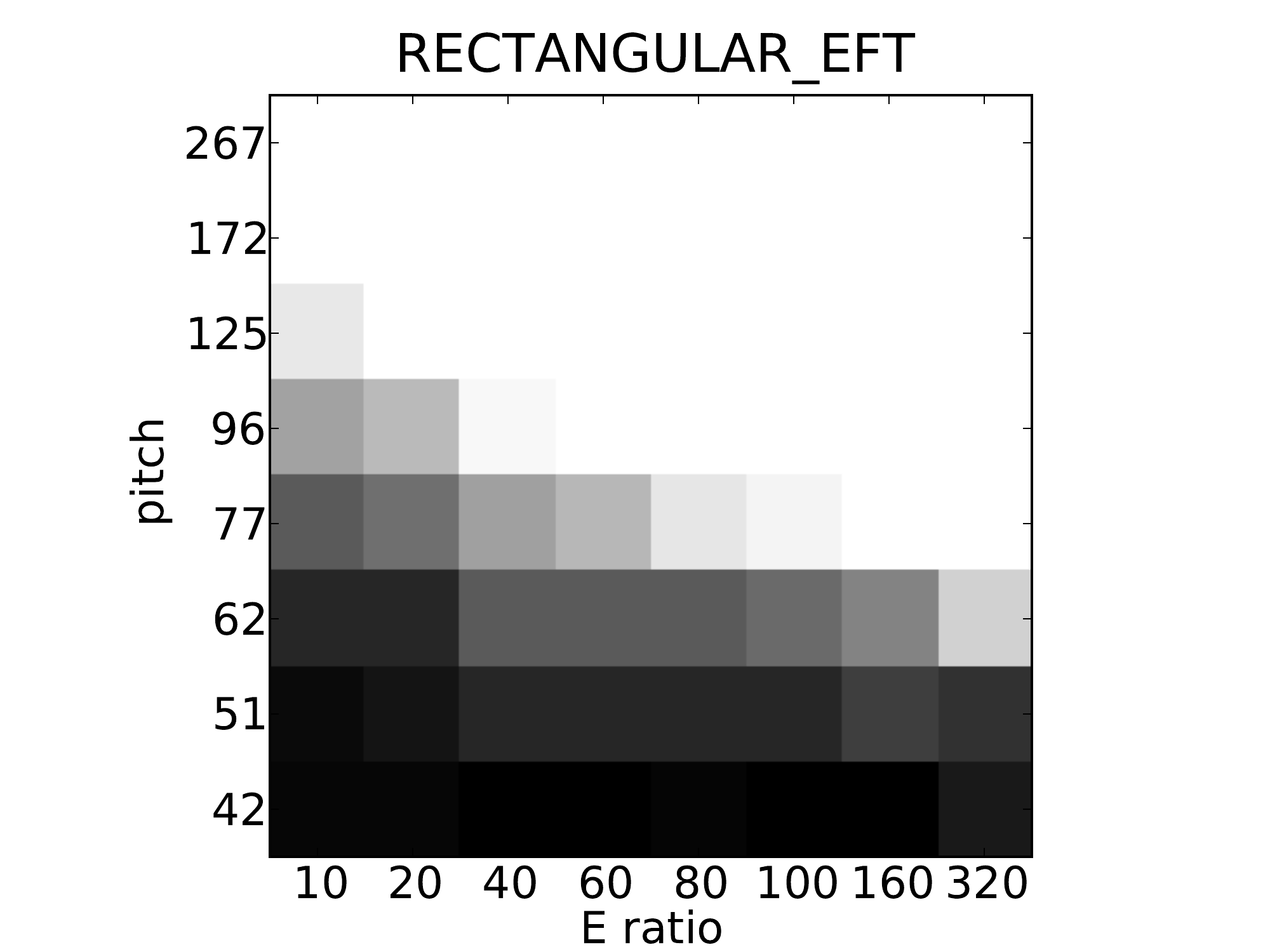}
\includegraphics[width=0.23\linewidth,clip=true]{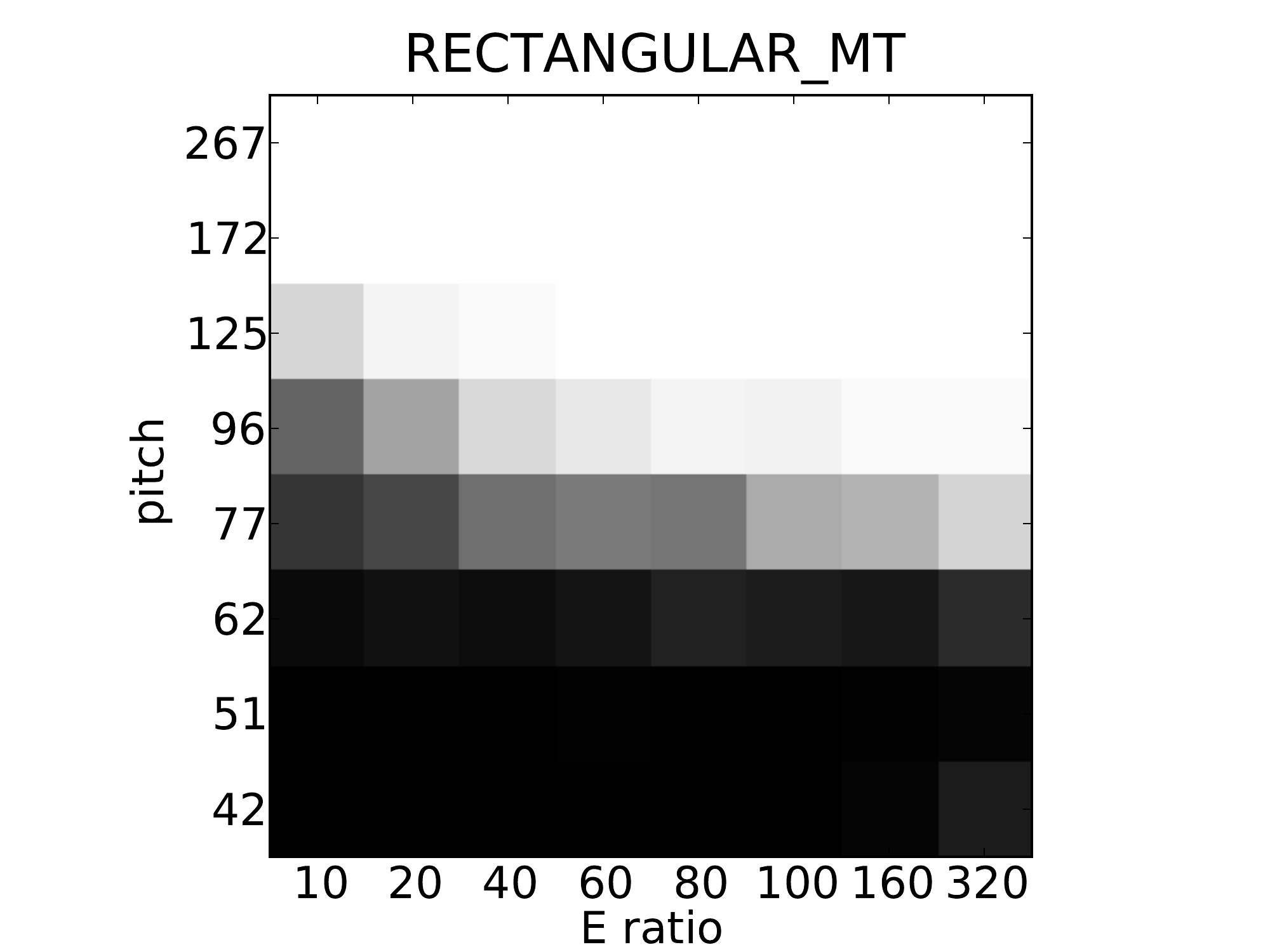}
\includegraphics[width=0.23\linewidth,clip=true]{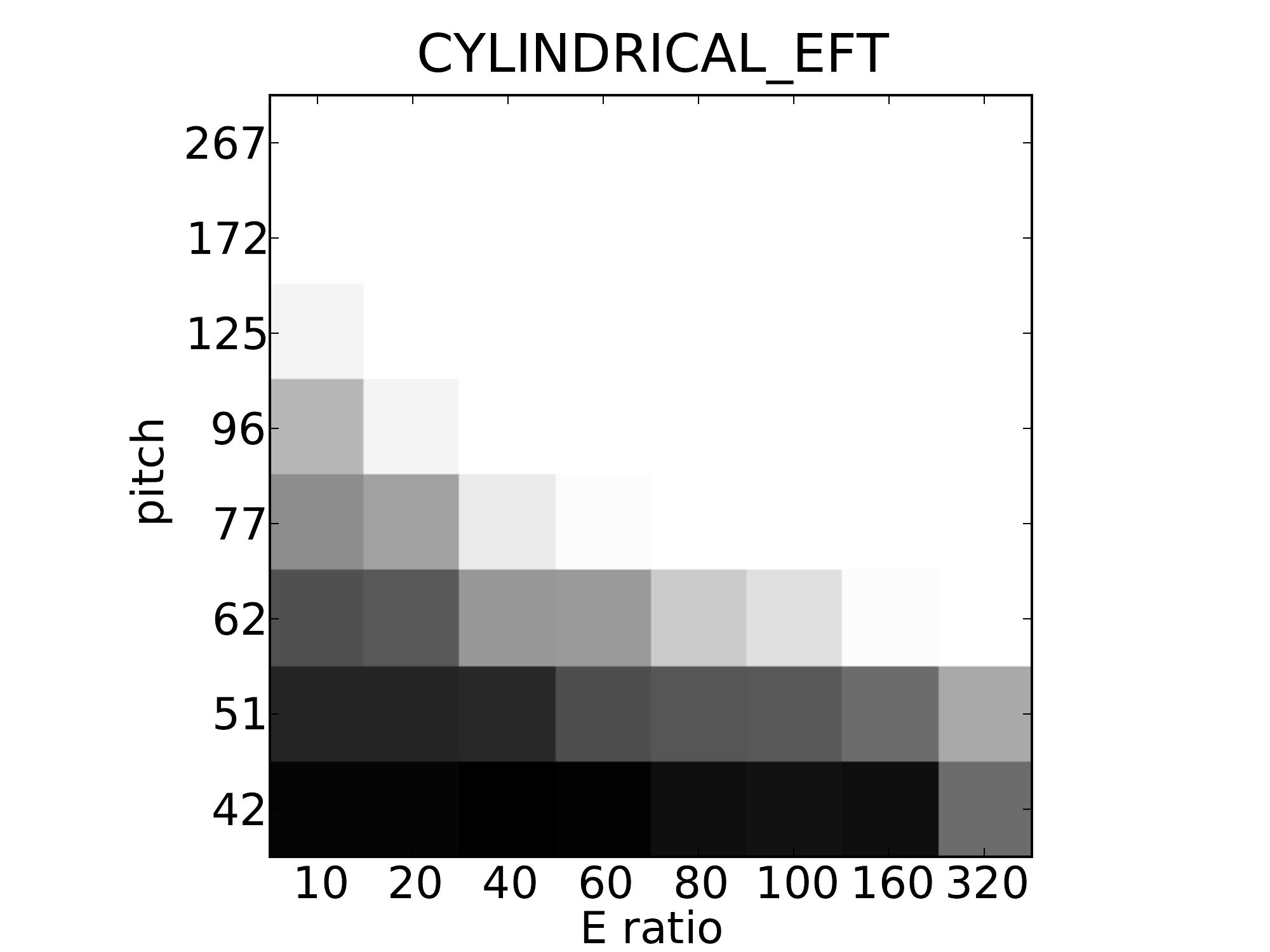}
\includegraphics[width=0.23\linewidth,clip=true]{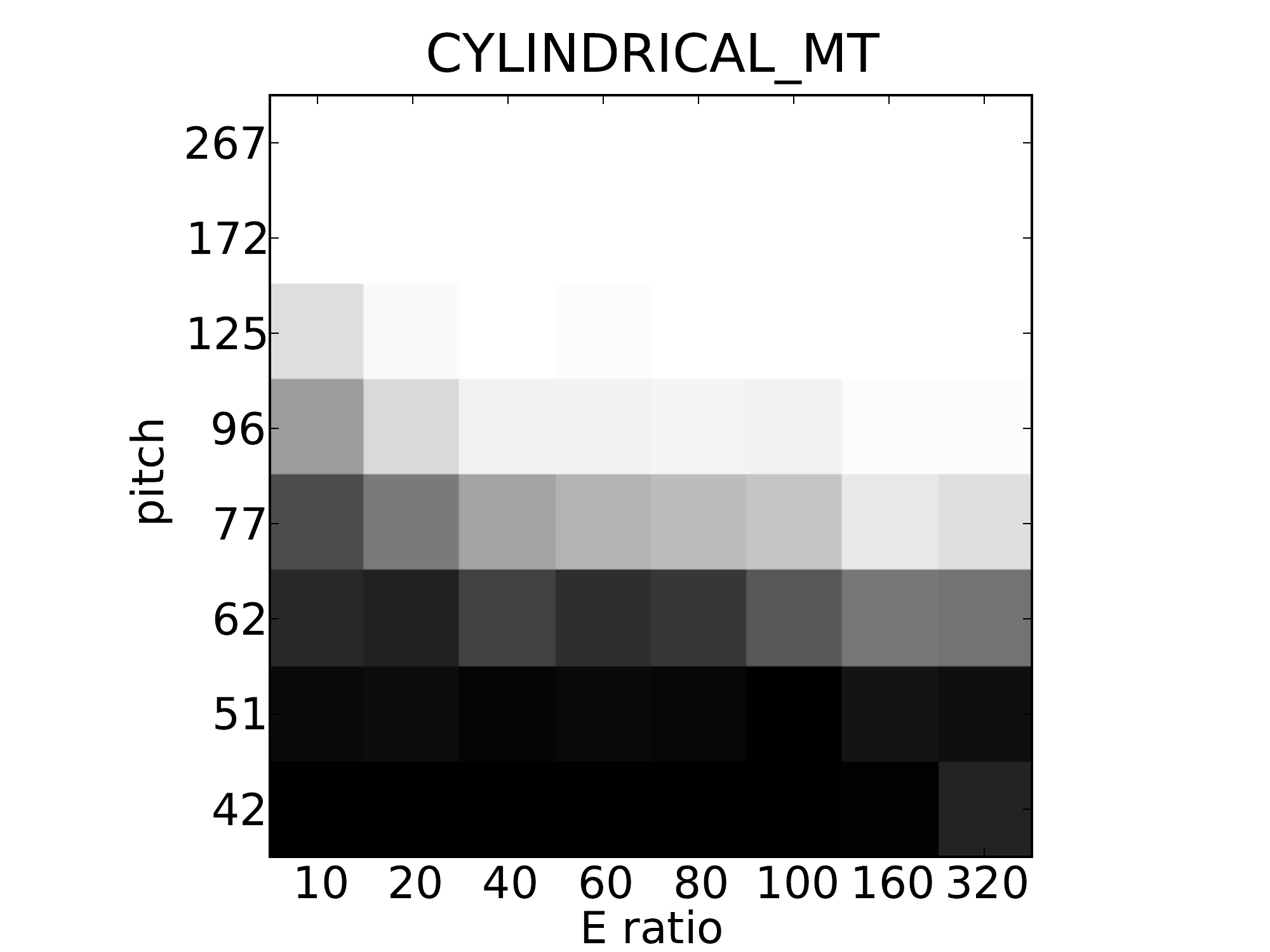}
\includegraphics[width=0.23\linewidth,clip=true]{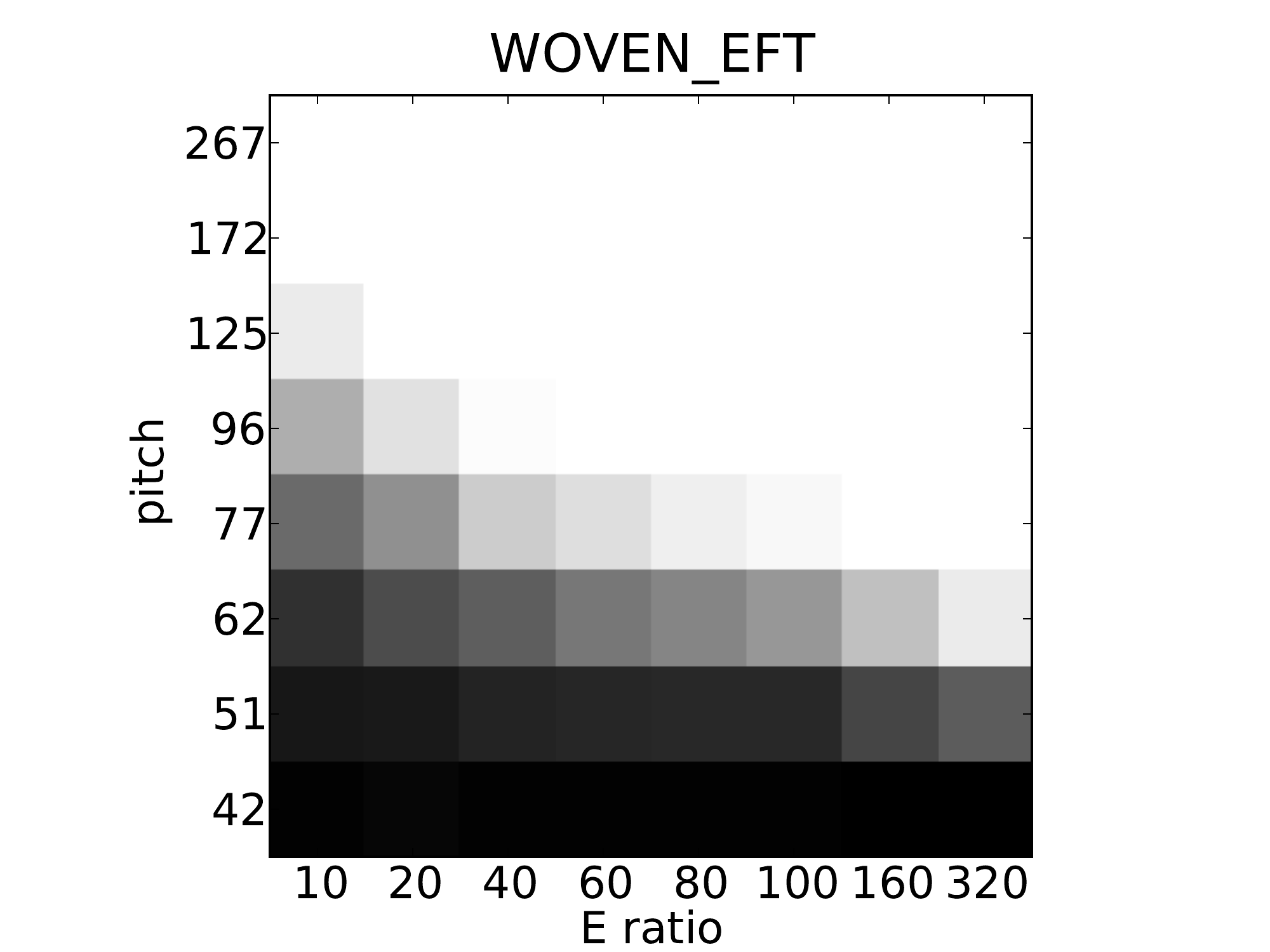}
\includegraphics[width=0.23\linewidth,clip=true]{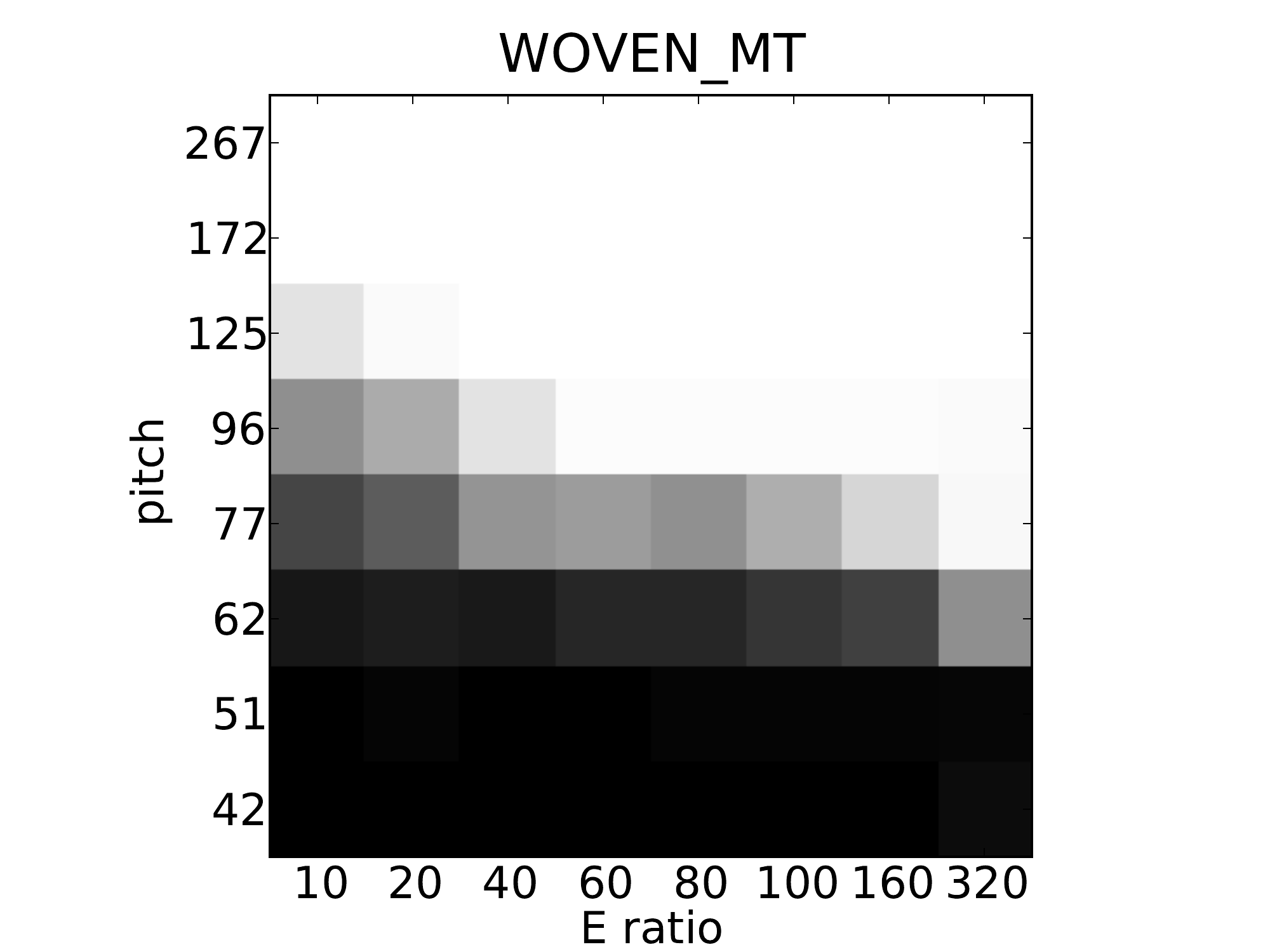}
\includegraphics[width=0.23\linewidth,clip=true]{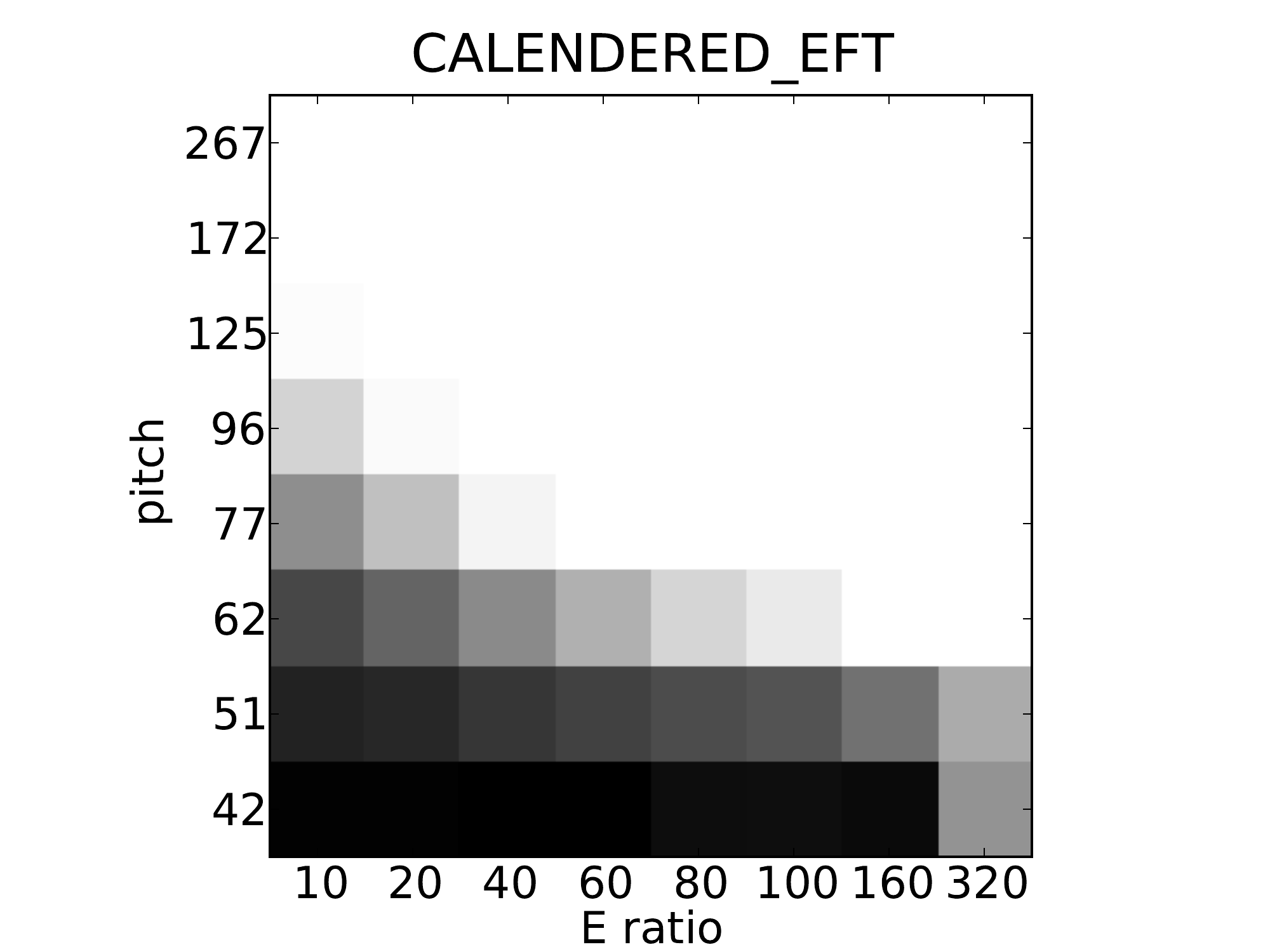}
\includegraphics[width=0.23\linewidth,clip=true]{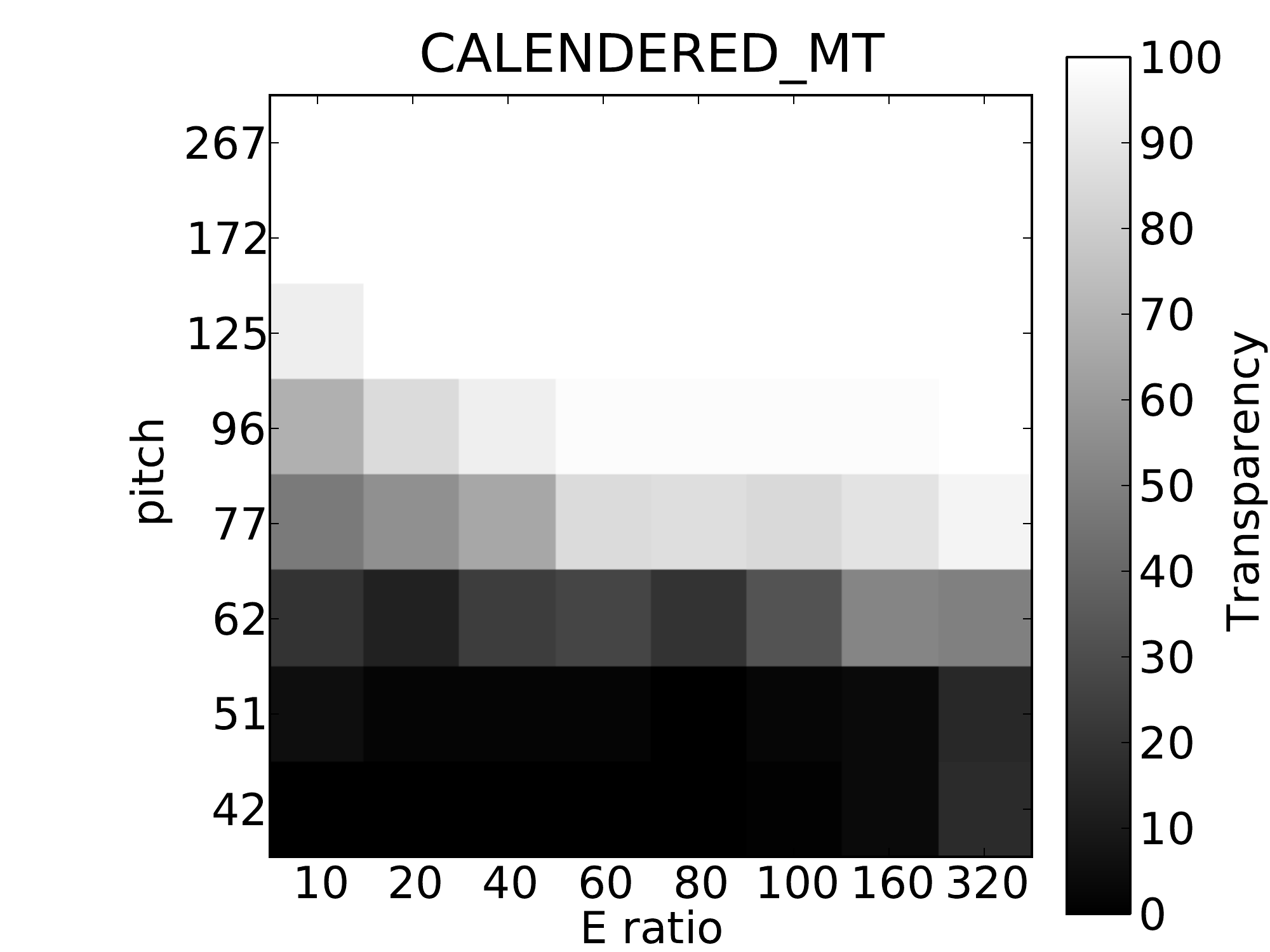}
}
\end{center}
\caption{Electrical transparencies calculated with EFT and MT methods for different types of meshes. Pitch is given in $\mu$m and with a wire diameter of d=28.2 $\mu$m the pitches and corresponding optical transparencies in percent are 42 (10\%), 51 (20\%), 62 (30\%), 77 (40\%), 96 (50\%), 125 (60\%), 172 (70\%) and 267 (80\%). }
\label{electrical_transparency}
\end{figure*}

In a particular case of a fixed mesh parameter (optical transparency of 40\%) electrical transparency increases with the increase of the amplification to drift field ratio. In the MT calculations, the cylindrical, woven and calendered mesh are very similar while rectangular mesh yields much lower transparencies than the other three - (see figure \ref{electrical_transparency}). 
Our explanation of this behavior lies in the shape of the deflection electric field (i.e. electric field in the x-y direction), parallel to the mesh, which deflects electrons from hitting the mesh and guides them into the holes yielding the funnel effect shown in figure \ref{lines_of_force}. 
Electric field values in the x-y plane are shown in figure \ref{cross_xy_fld_pot} for all 4 types of meshes. The cause of the weakest funneling power of the rectangular mesh lies in its high symmetry which causes large electric field cancellation in the x-y direction. This effect is to some extent present in the cylindrical mesh and in the calendered and woven mesh one can nicely see the reduction in symmetry of the geometry of the mesh which is reflected in the shape of the $E_{xy}$ electric field.\\

The result for the gain, given in figure \ref{gain_at_pitch_77}, is very interesting because it shows that 4 meshes are practically indistinguishable in this respect. The explanation for this is rather simple. The gain is obtained as an integral of the Townsend coefficient along the trajectory of the particle that goes from drift region into amplification region and depends on the electric field values along this path. All 4 meshes (or virtually any other shape of the mesh) generate a homogeneous field above and below the mesh in the large portion of the volume, and differences in the electric field are visible only very close to the wires of the mesh. 
 Electrons that pass near wires account for a very small fraction of all trajectories (confirm Fig. 3), and therefore the mesh type has no significant influence on the gain.
Hence the crucial role of the geometry of the mesh is whether the electron makes it from the drift into amplification region without bumping into the mesh which determines the electrical transparency of the mesh and is obviously very sensitive to those electric field details around the mesh. On the other hand the electrons that go through achieve the gain that is almost completely determined just by the ratio of the amplification and drift field and is not sensitive to the detailed geometry.

\begin{figure*}[htb]
\begin{center}
{
\includegraphics[width=0.9\linewidth,clip=true,angle=0]{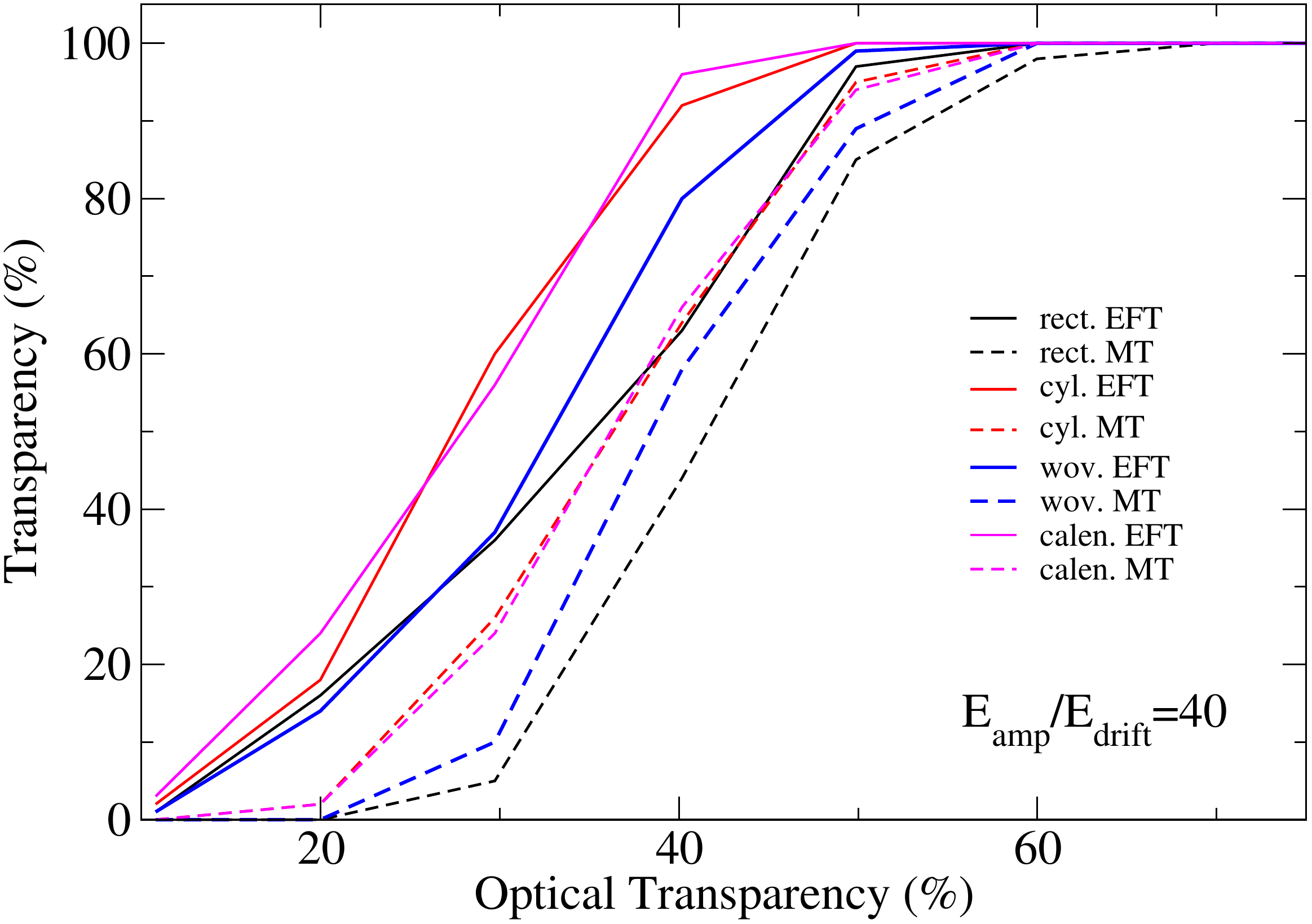}
}
\end{center}
\caption{Electrical vs. optical transparency. For relation between mesh pitch and optical transparency see caption of figure \protect \ref{electrical_transparency}. Full lines show calculation with EFT method while dashed lines represent results of MT calculations. In all cases amplification to drift field ratio is 40. }
\label{transparency_field_ratio_40}
\end{figure*}

\begin{figure*}[htb]
\begin{center}
{
\includegraphics[width=0.9\linewidth,clip=true]{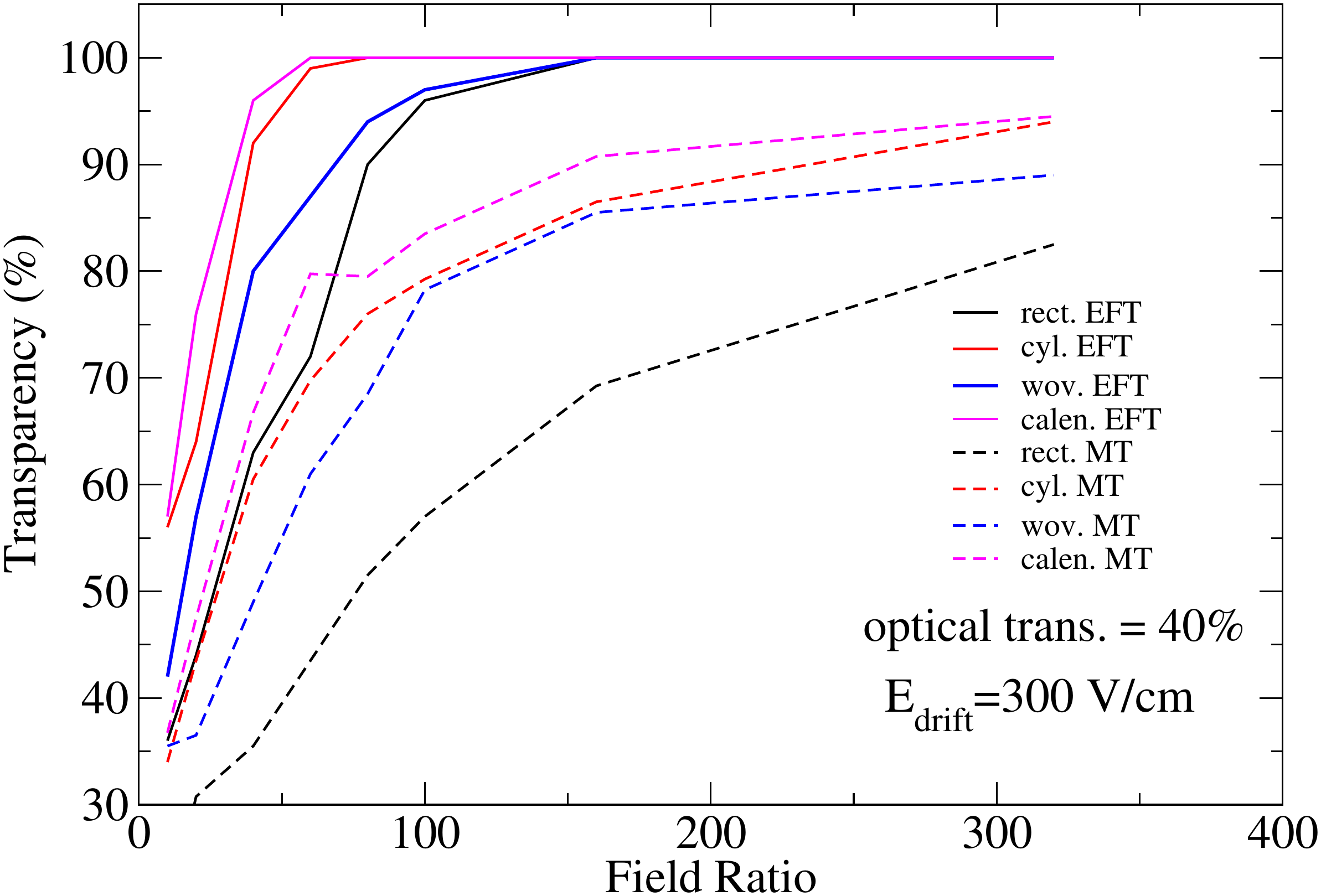}
}
\end{center}
\caption{Electrical transparency as a function of an amplification to drift field ratio. All four meshes have identical pitch of 77 $\mu$m which is equivalent to 40\% optical transparency.}
\label{transparency_at_pitch_77_40_percent_optical}
\end{figure*}

\begin{figure*}[htb]
\begin{center}
{
\includegraphics[width=0.9\linewidth,clip=true]{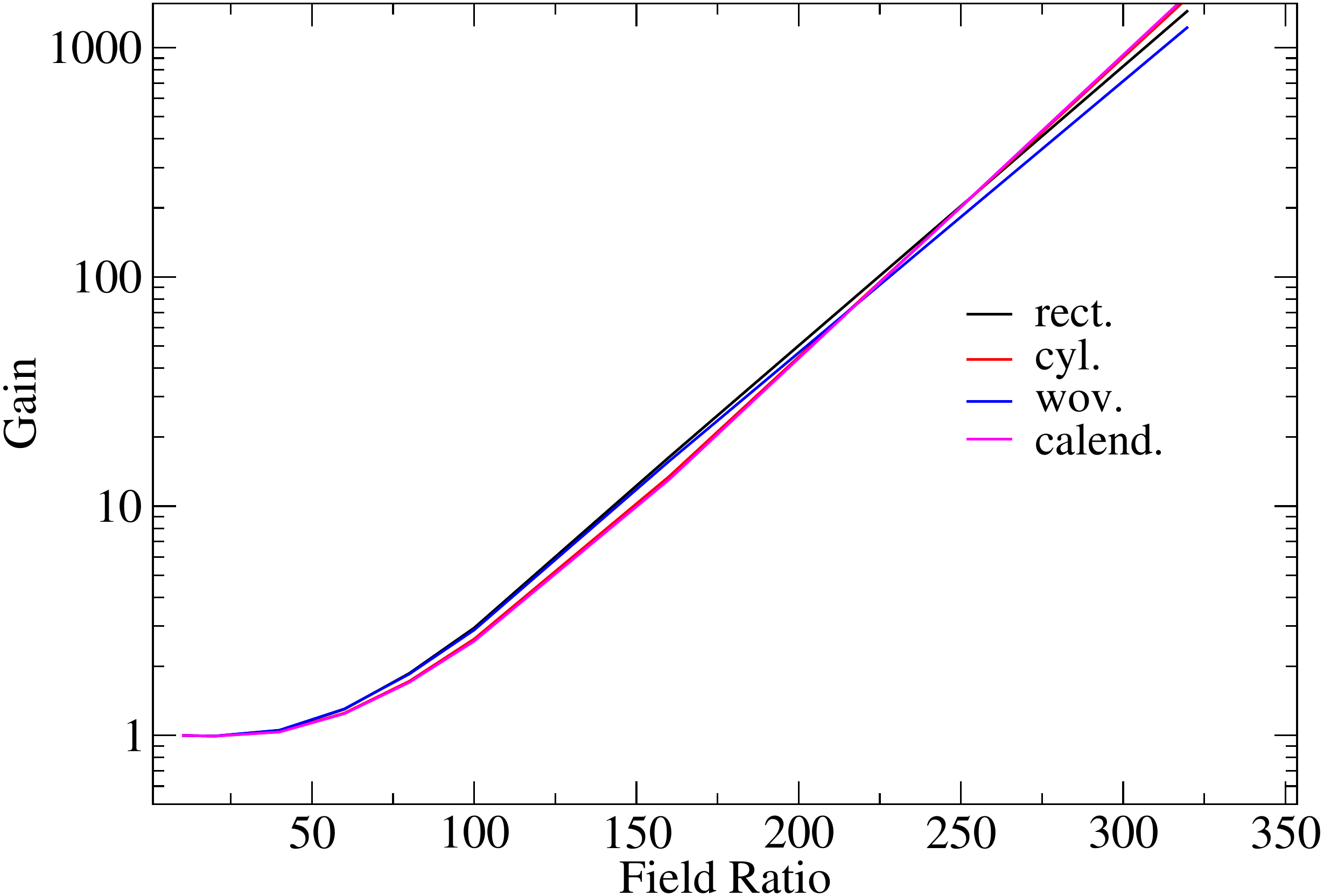}
}
\end{center}
\caption{Gain as a function of an amplification to drift field ratio.} 
\label{gain_at_pitch_77}
\end{figure*}

\begin{figure*}[htb]
\begin{center}
{
\includegraphics[width=0.45\linewidth,clip=true]{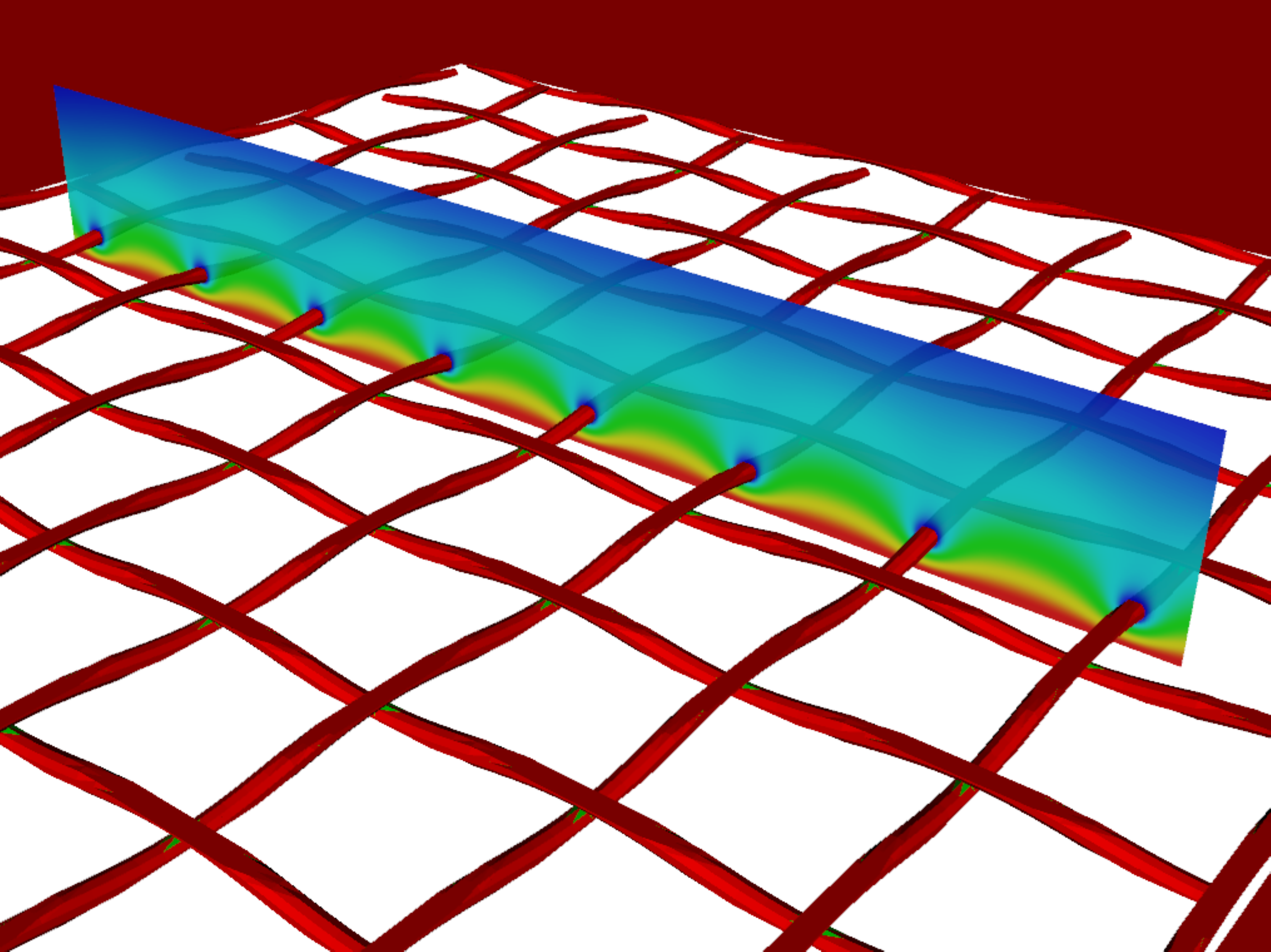}
\includegraphics[width=0.45\linewidth,clip=true]{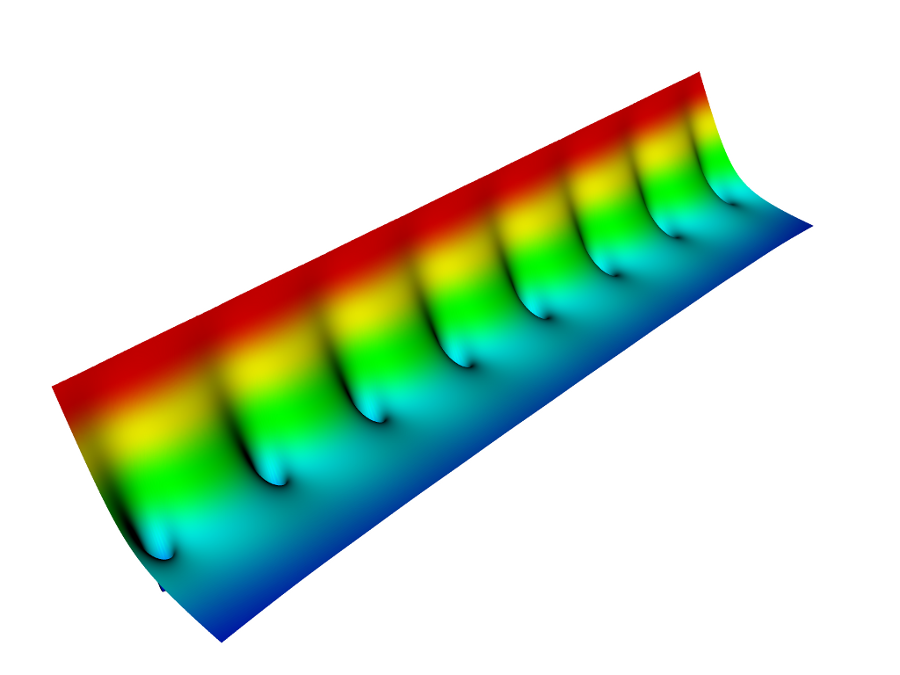}
\includegraphics[width=0.45\linewidth,clip=true]{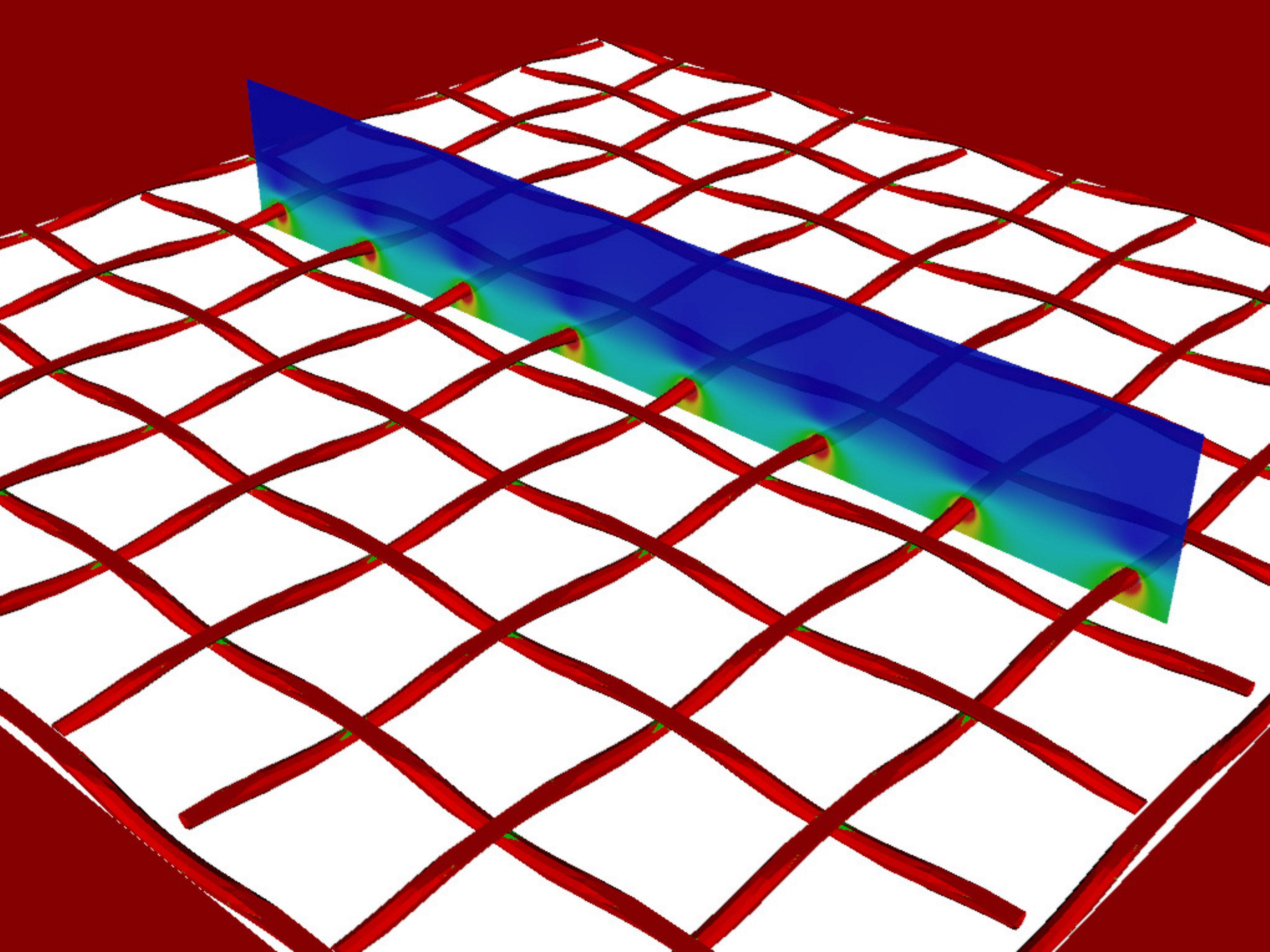}
\includegraphics[width=0.45\linewidth,clip=true]{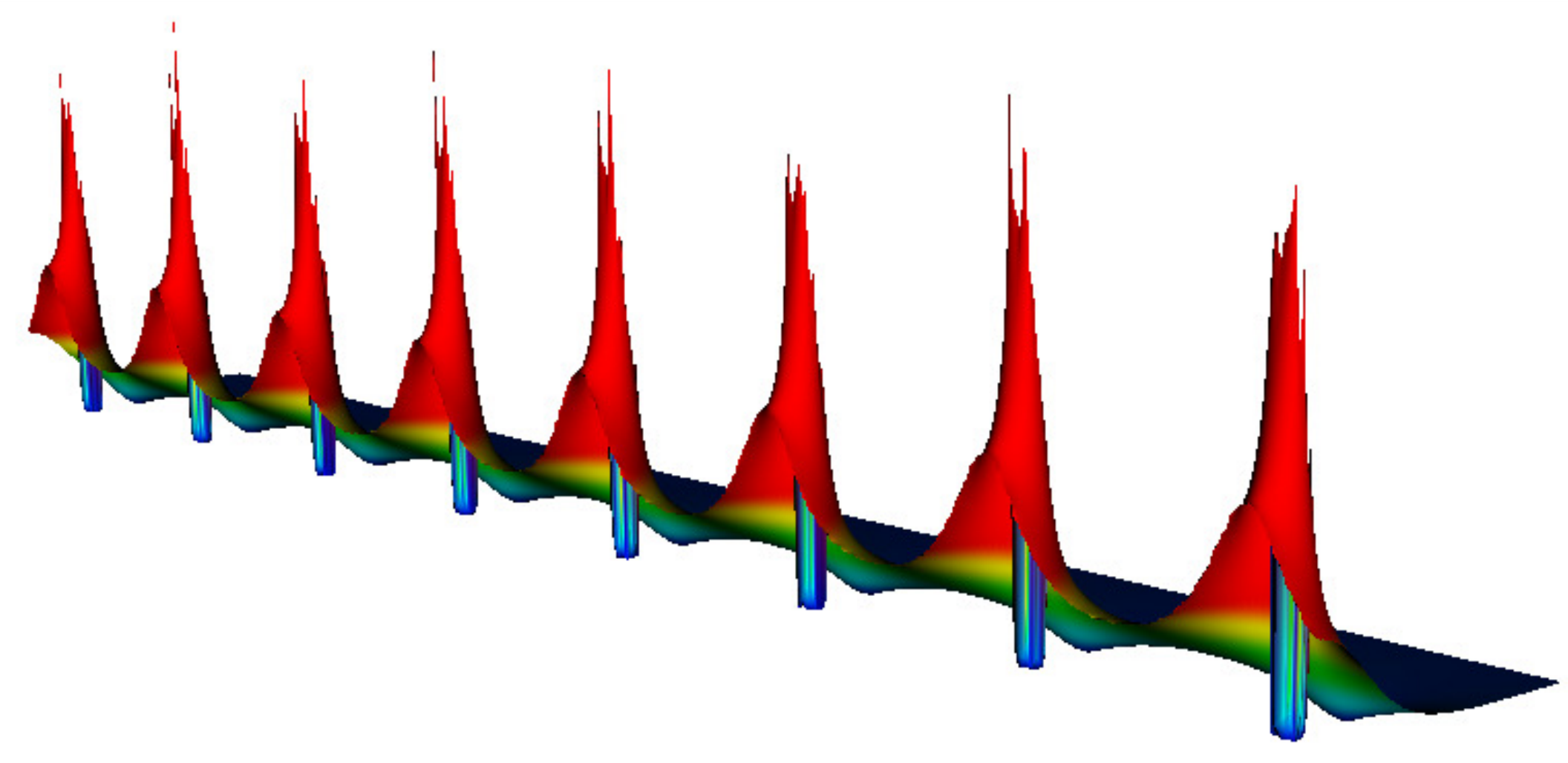}
}
\end{center}
\caption{Top row: Electric potential shown in a cross section plane - shown on left. Right - the same cross section plane with potential value protruded in the direction perpendicular to the plane. 
Bottom row: Electric field shown in a cross section plane - shown on left. Right - the cross section plane protruded in perpendicular direction proportionally to the field value. Notice that inside the wire (mesh) electric field drops exactly to zero as expected from the Faraday's cage effect - this, among other things, shows incredible accuracy of the BEM solutions in electrostatic.
}
\label{potential}
\end{figure*}


\begin{figure*}[htb]
\begin{center}
{
\includegraphics[width=0.9\linewidth,clip=true]{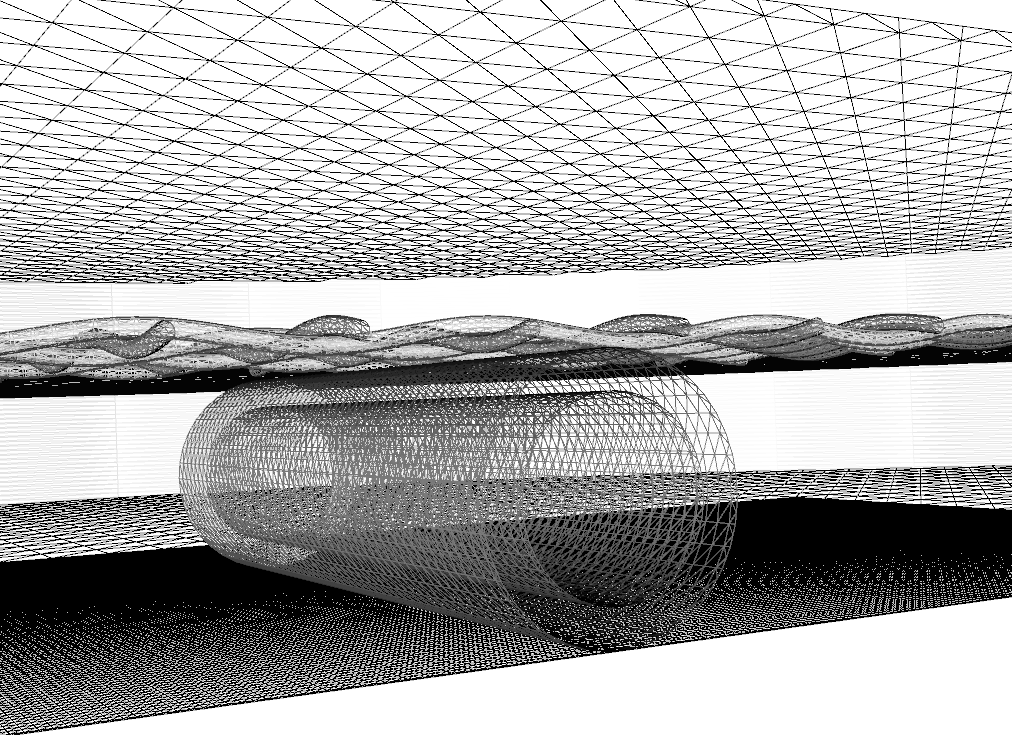}
}
\end{center}
\caption{Geometry of the dielectric spacer placed between bottom electrode and the mesh. In this figure hollow spacer geometry is shown. Outer diameter of the spacer is 450 $\mu$m while inner diameter is 320 $\mu$m.}
\label{hollow_spacer}
\end{figure*}

\begin{figure*}[htb]
\begin{center}
{
\includegraphics[width=0.32\linewidth,clip=true]{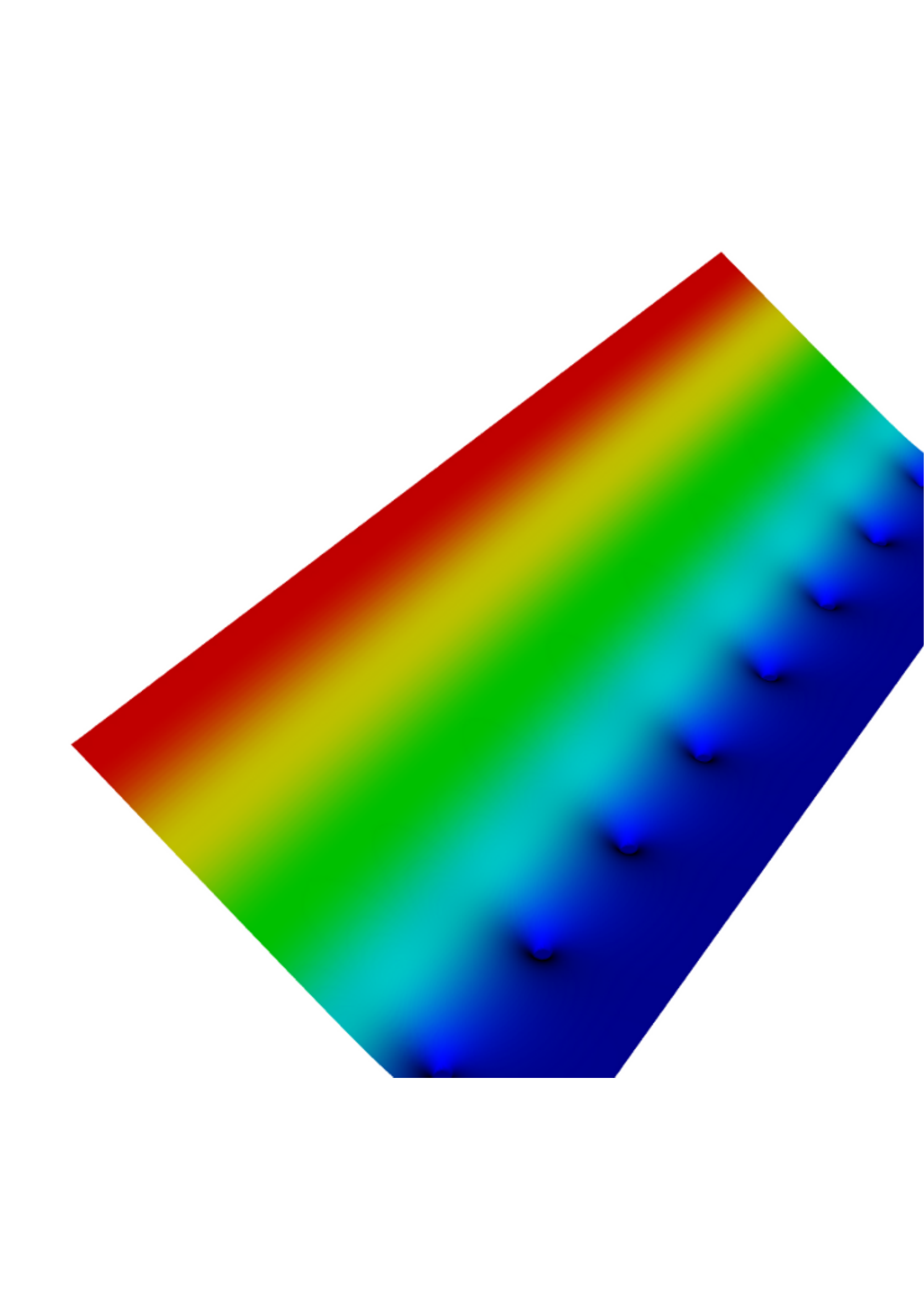}
\includegraphics[width=0.32\linewidth,clip=true]{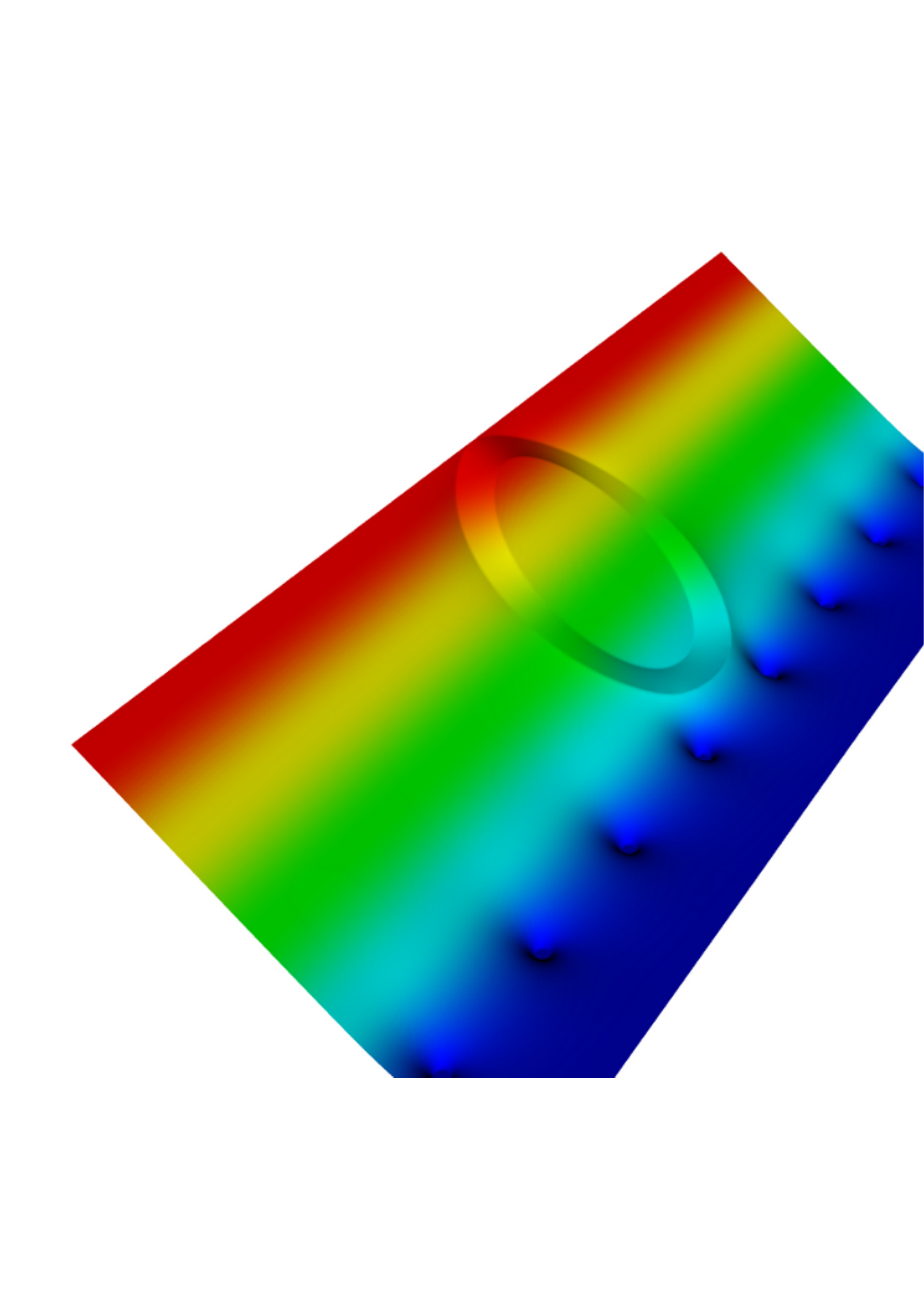}
\includegraphics[width=0.32\linewidth,clip=true]{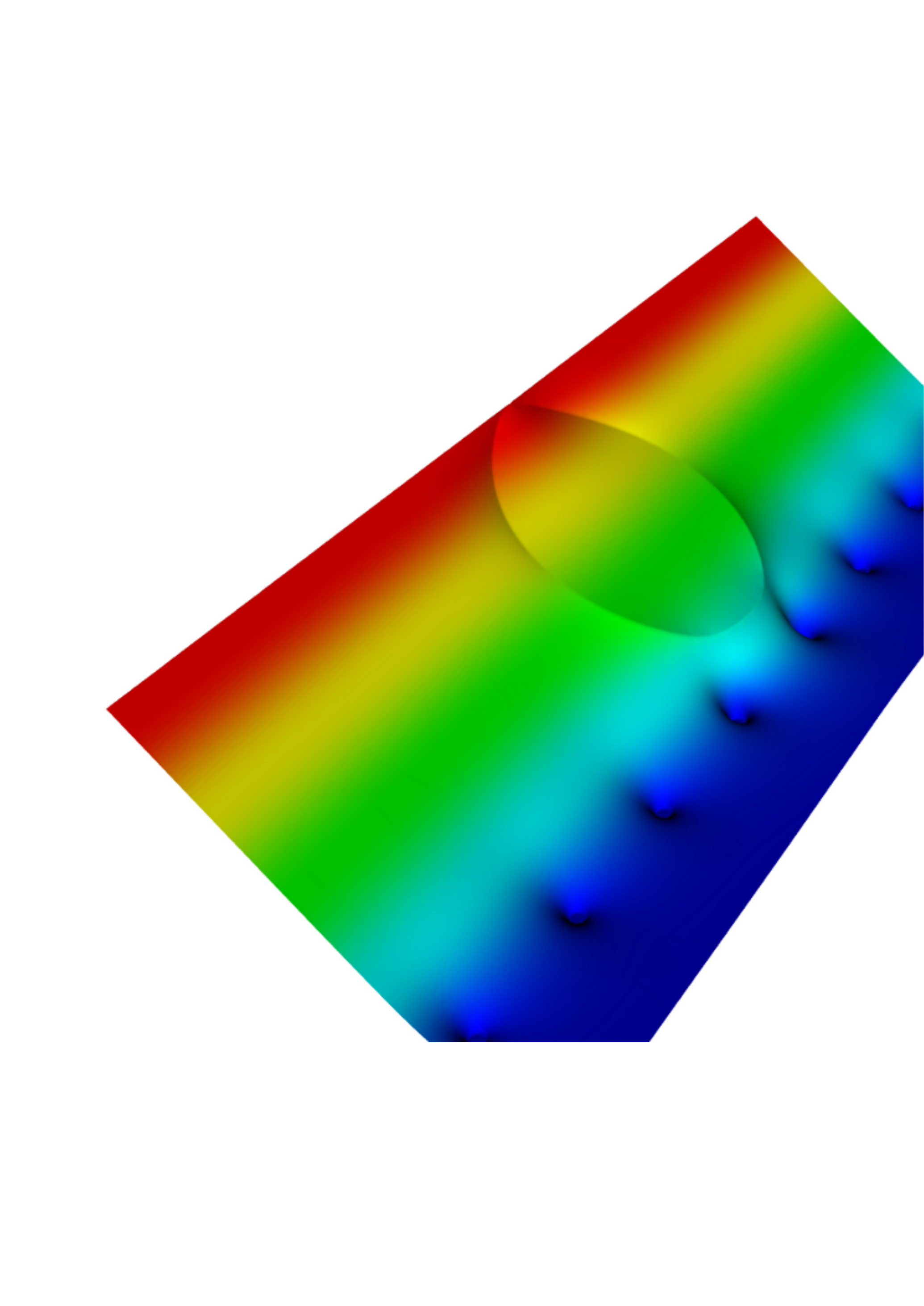}
\includegraphics[width=0.32\linewidth,clip=true]{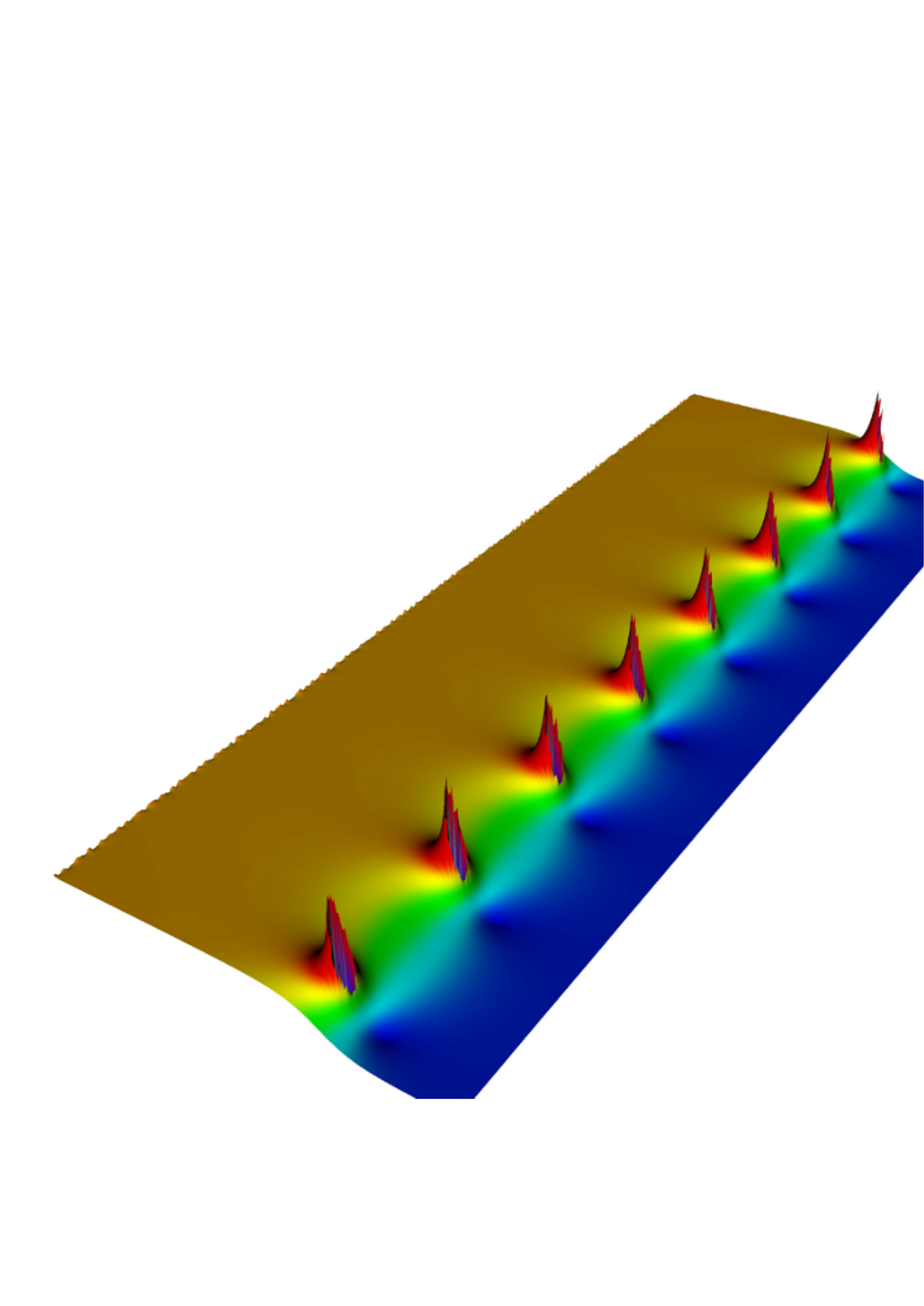}
\includegraphics[width=0.32\linewidth,clip=true]{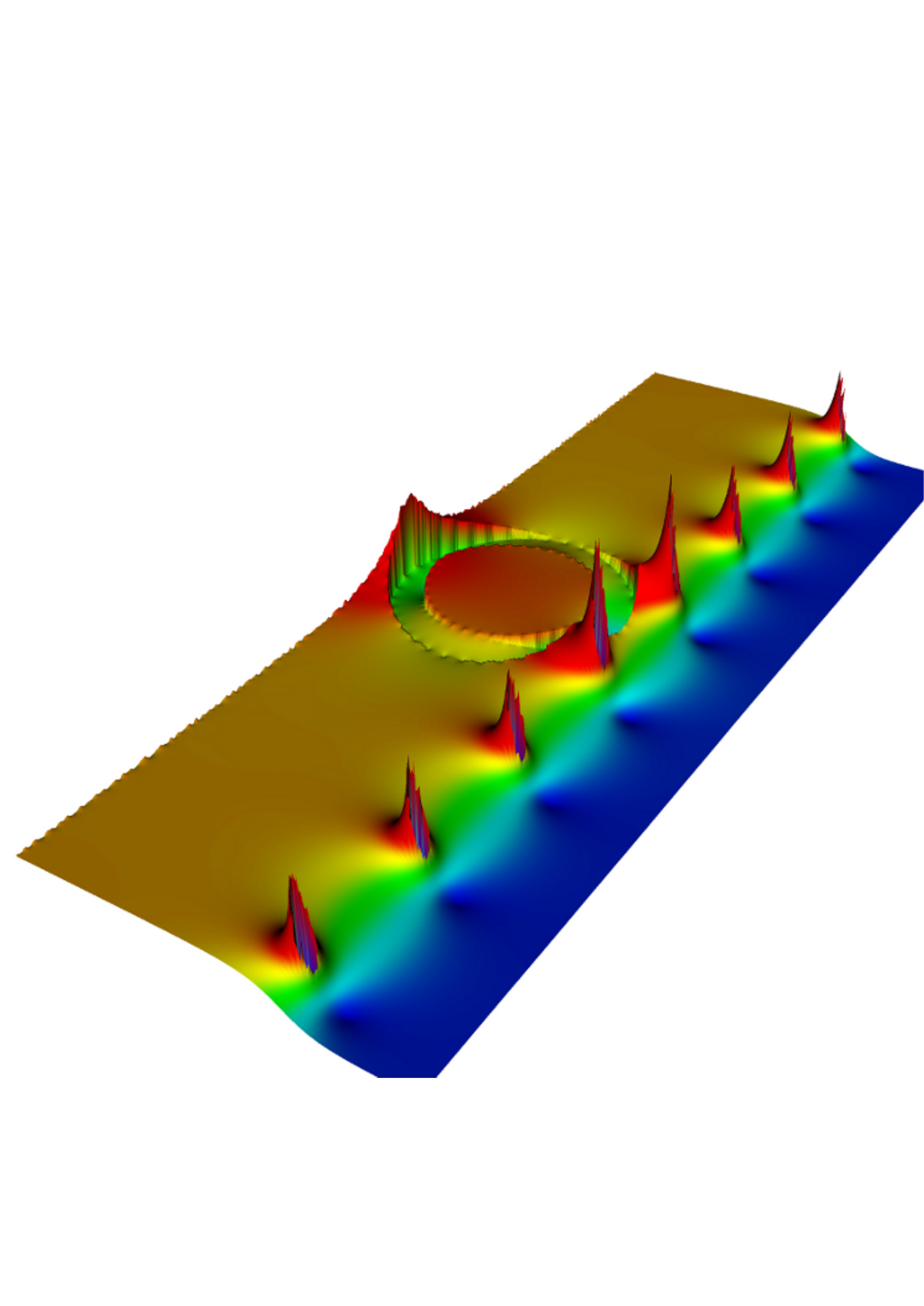}
\includegraphics[width=0.32\linewidth,clip=true]{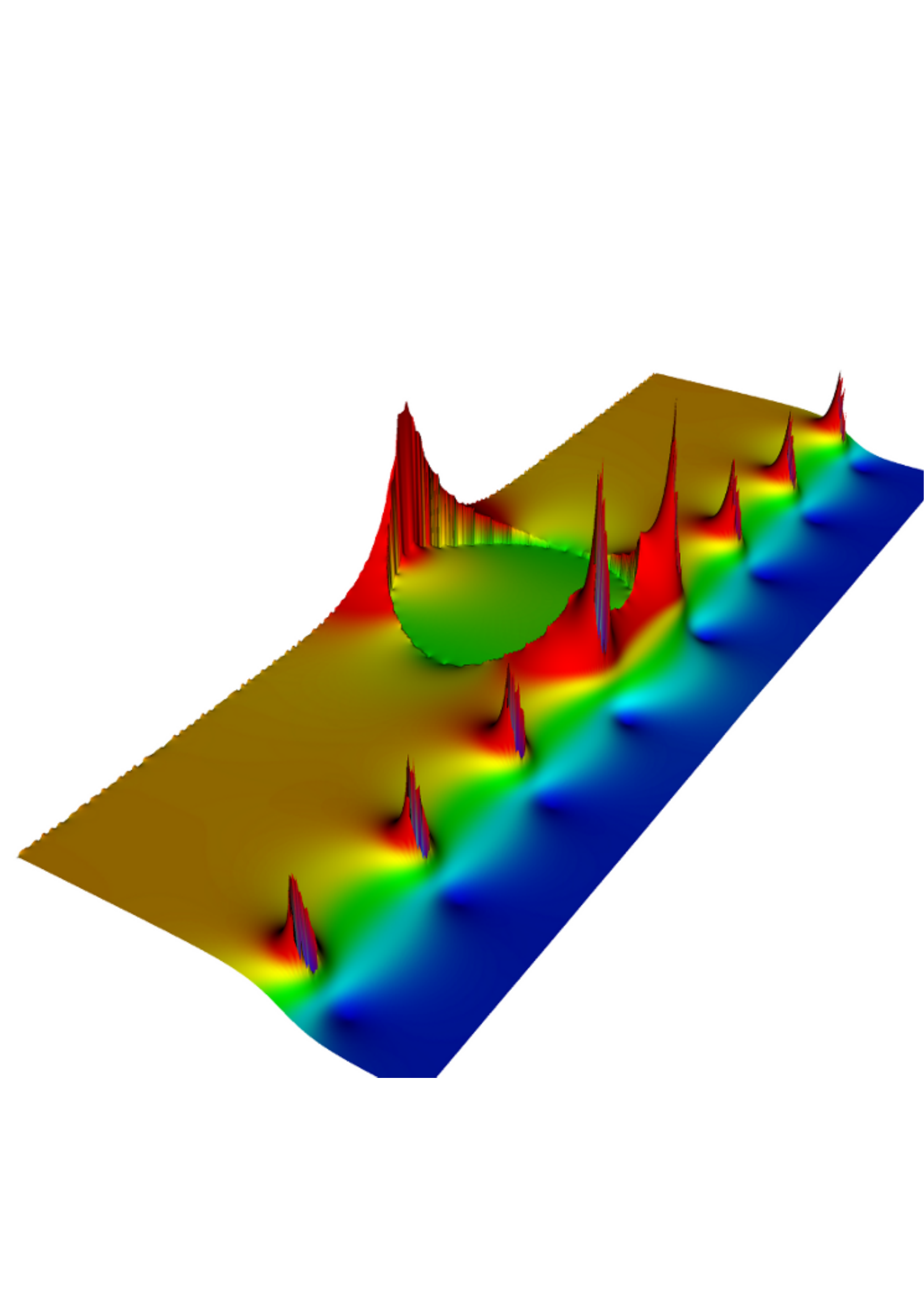}
}
\end{center}
\caption{Electric potential (top) and field (bottom) along the plane cut shown in figure \protect \ref{potential}. Left - without spacer. Middle - with hollow (capillary) spacer. Right - with full cylindrical spacer. Comparing maximum electric field values for all three cases gives larger maximum fields for the cases with spacer present. Namely, in the case with hollow spacer maximum field is 1.4 times larger than in the case without a spacer while in the case with the full spacer this factor is 2.0. 
}
\label{spacer_impact_potential}
\end{figure*}

\begin{figure*}[htb]
\begin{center}
{
\includegraphics[width=0.32\linewidth,clip=true]{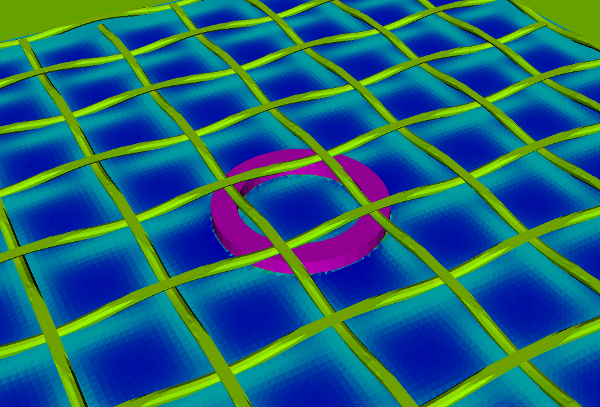}
\includegraphics[width=0.32\linewidth,clip=true]{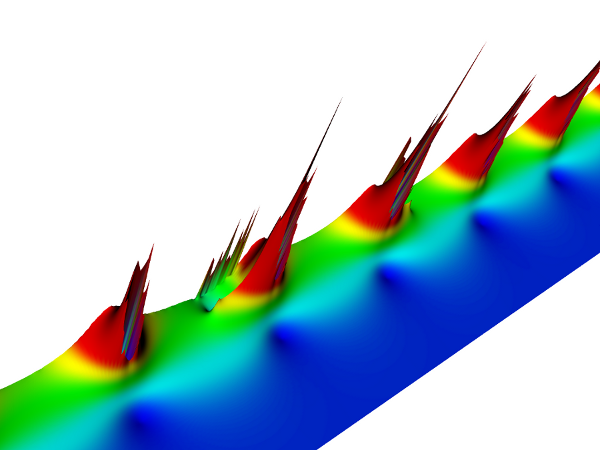}
\includegraphics[width=0.32\linewidth,clip=true]{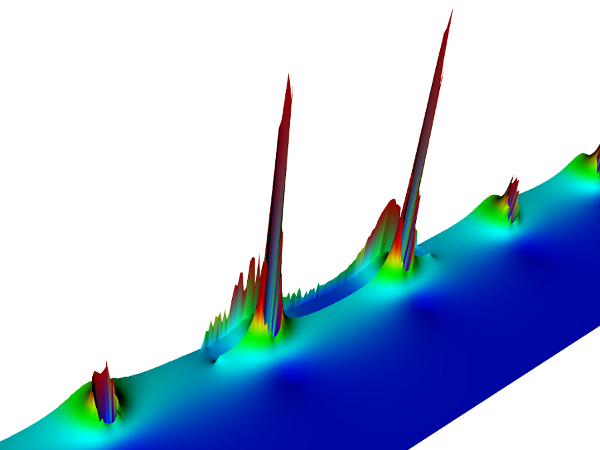}
}
\end{center}
\caption{Left - geometry of the vertical hollow spacer. Middle - electric field in the vertical cross section in system with hollow spacer. Right - electric field in the vertical cross section in system with full spacer. Inner and outer diameter of the spacer are 320 $\mu$m and 450 $\mu$m respectively.
The height of the spacer is 37.8 $\mu$m.
A dielectric spacer causes increase of the electric field compared to a case wihthout a spacer. Moreover, in the case of hollow spacer maximum field value is increased by factor 1.25 while in the case of full spacer maximum field is 3.3 times larger than in the case without the spacer.
}
\label{spacer_vertical}
\end{figure*}


\begin{figure*}[htb]
\begin{center}
{
\includegraphics[width=0.9\linewidth,clip=true]{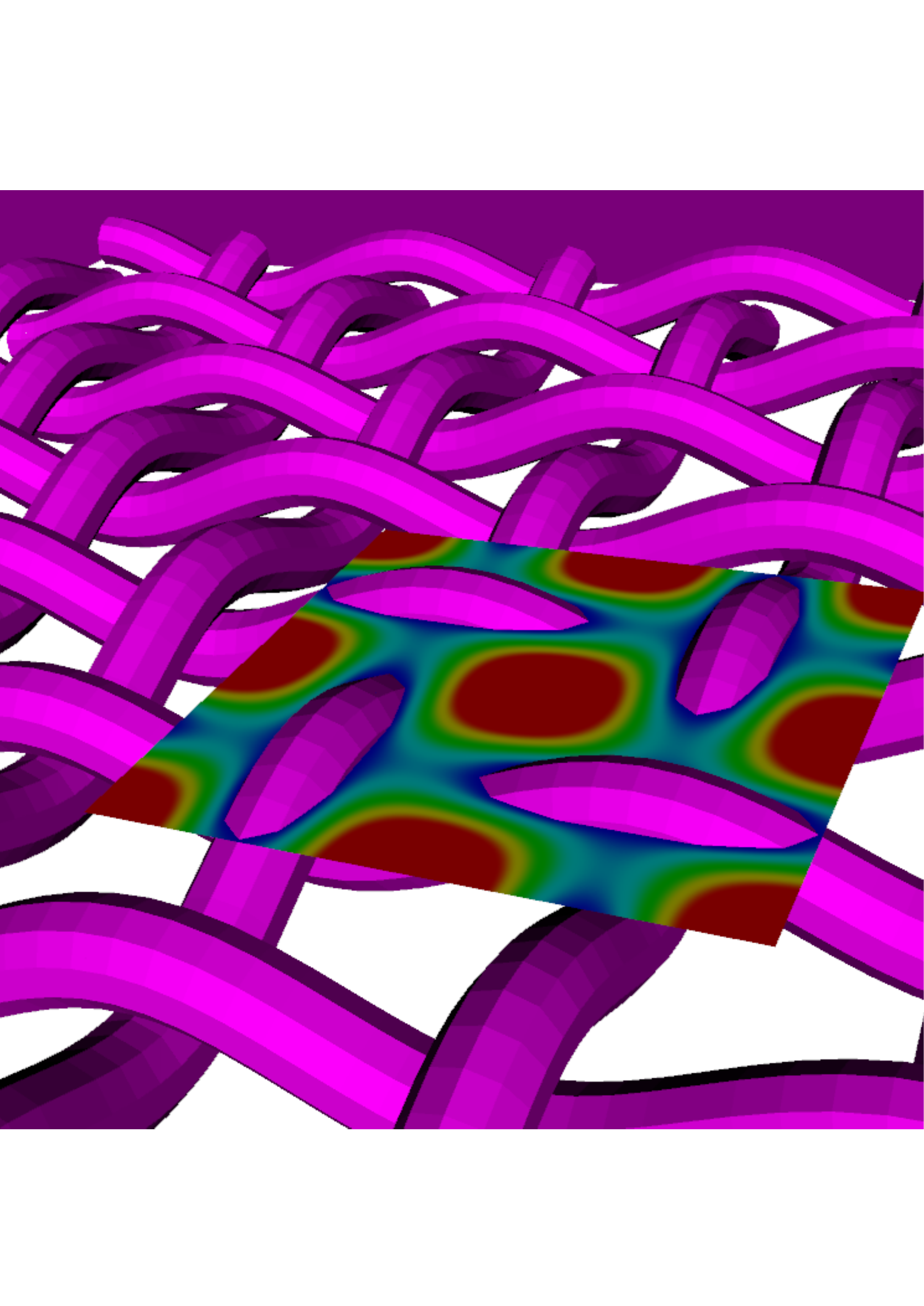}
}
\end{center}
\caption{Cross section region in a horizontal plane around one hole in the mesh. Detailed electric field values and potentials are shown in figure \protect \ref{cross_xy_fld_pot}.}
\label{cross_xy}
\end{figure*}

\begin{figure*}[htb]
\begin{center}
{
\includegraphics[width=0.23\linewidth,clip=true]{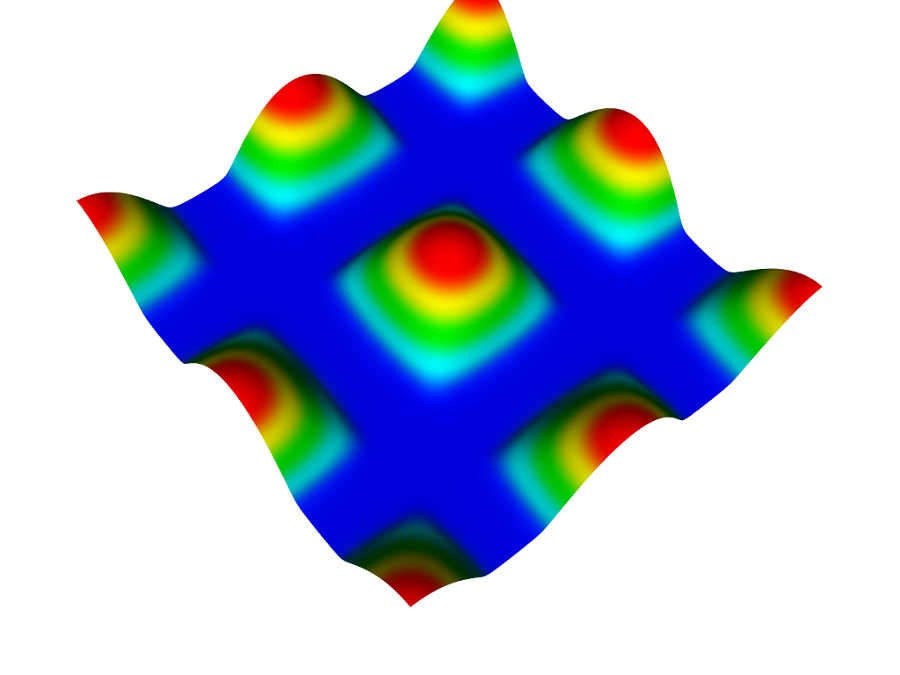}
\includegraphics[width=0.23\linewidth,clip=true]{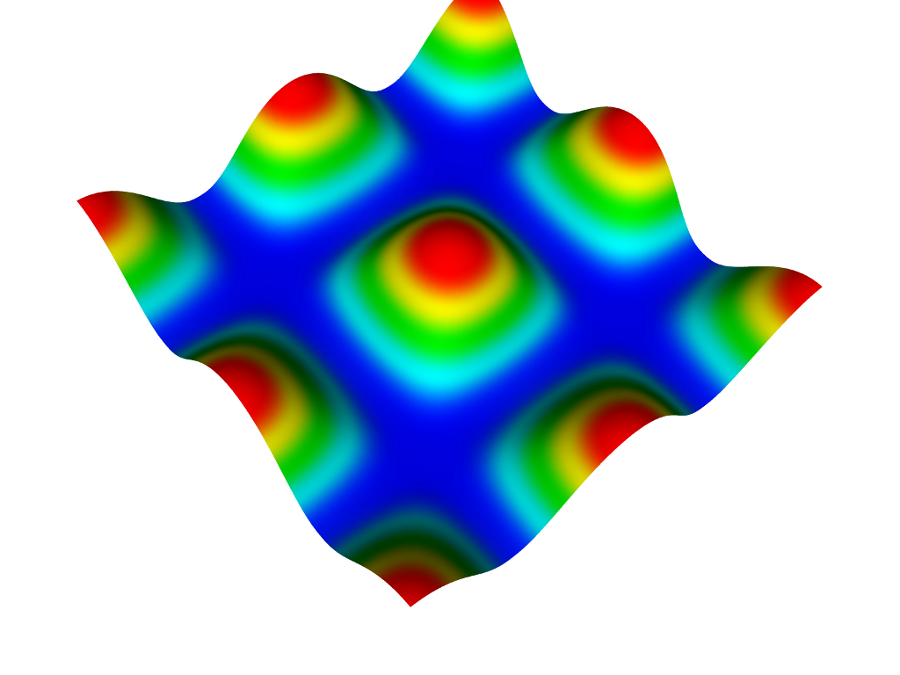}
\includegraphics[width=0.23\linewidth,clip=true]{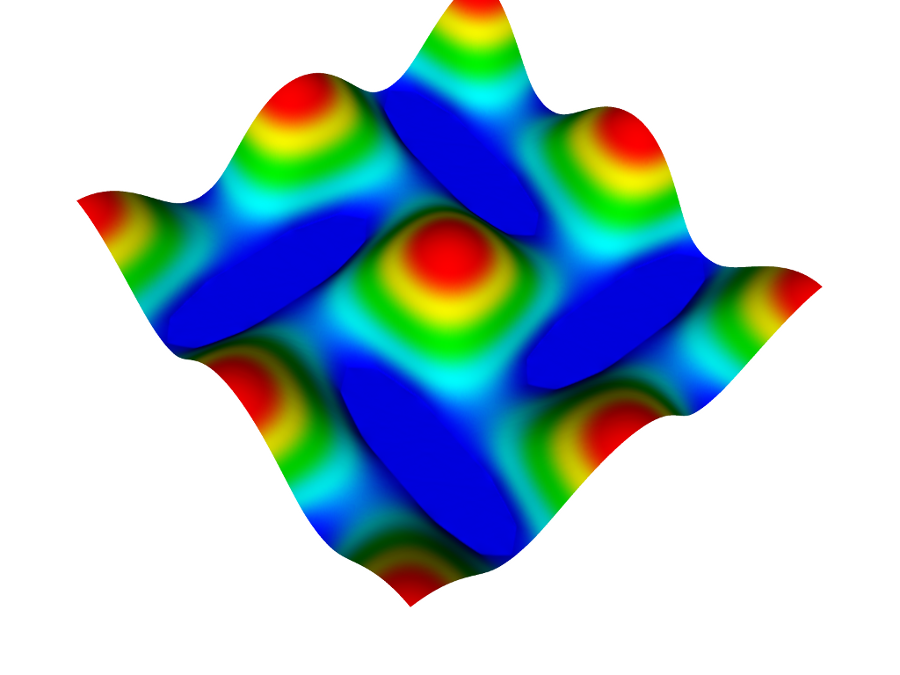}
\includegraphics[width=0.23\linewidth,clip=true]{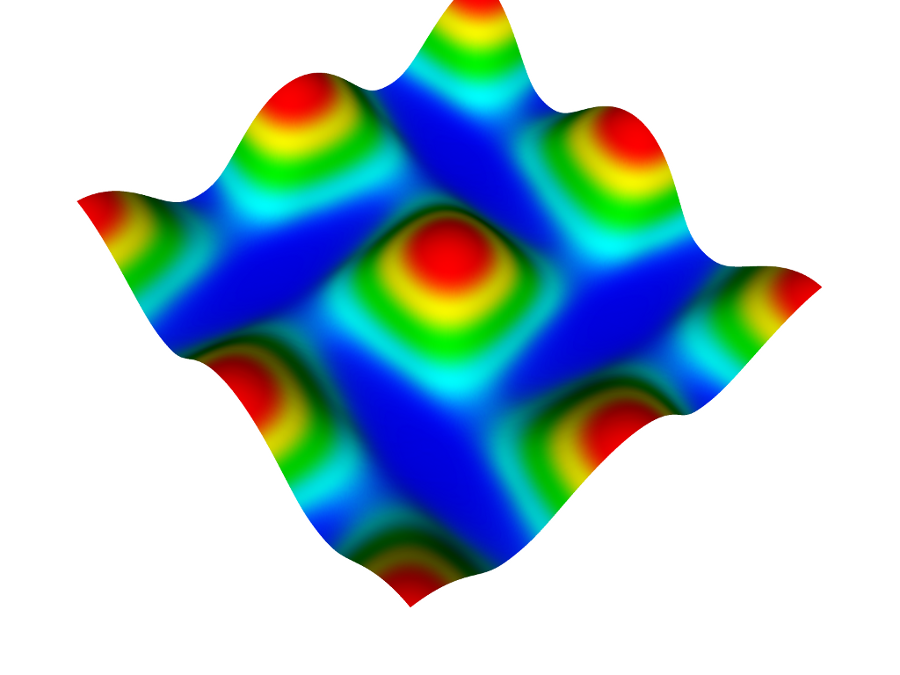}
\includegraphics[width=0.23\linewidth,clip=true]{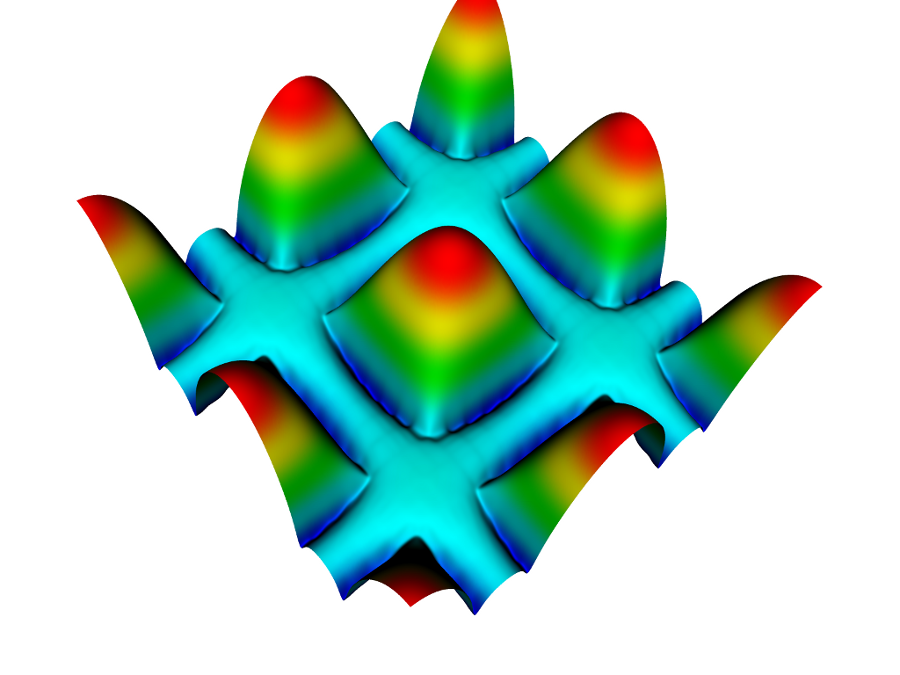}
\includegraphics[width=0.23\linewidth,clip=true]{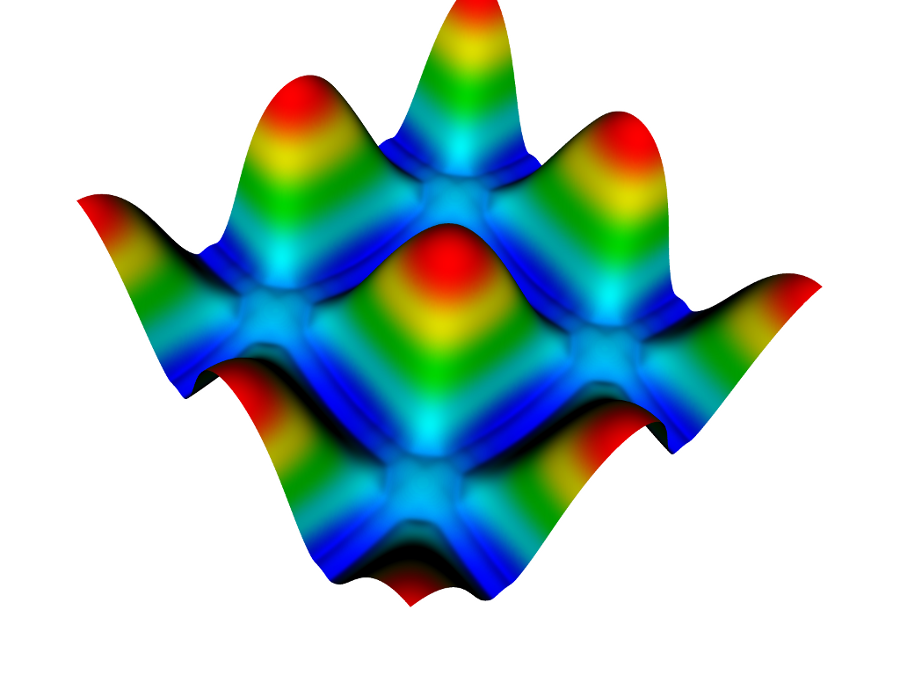}
\includegraphics[width=0.23\linewidth,clip=true]{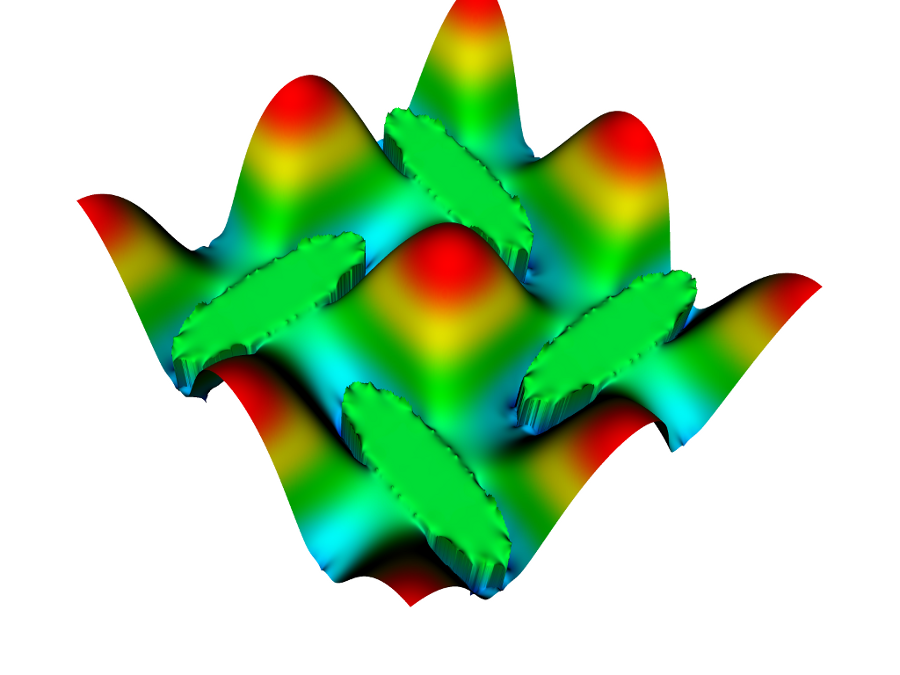}
\includegraphics[width=0.23\linewidth,clip=true]{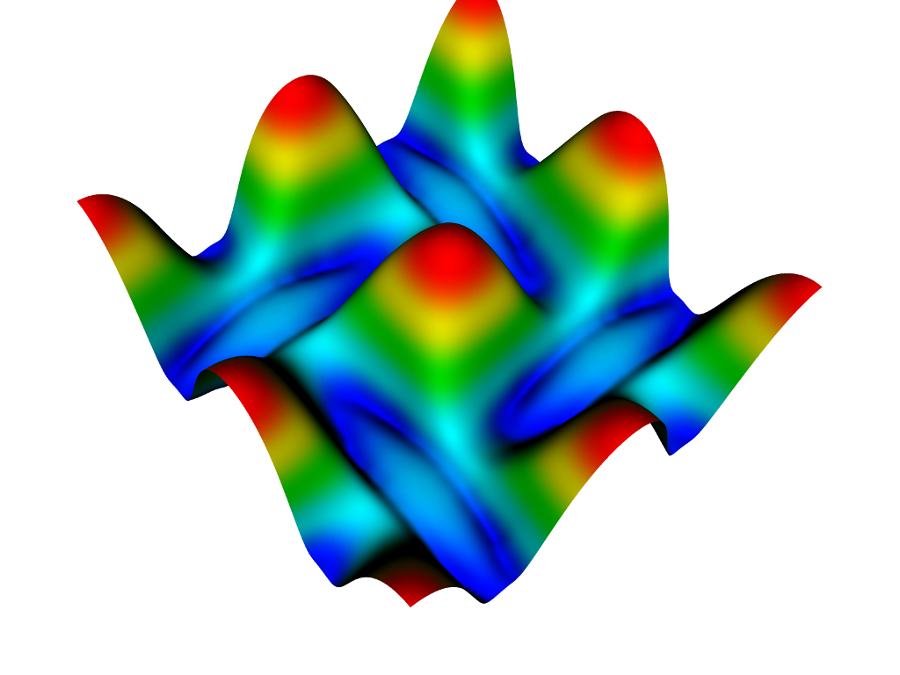}
\includegraphics[width=0.23\linewidth,clip=true]{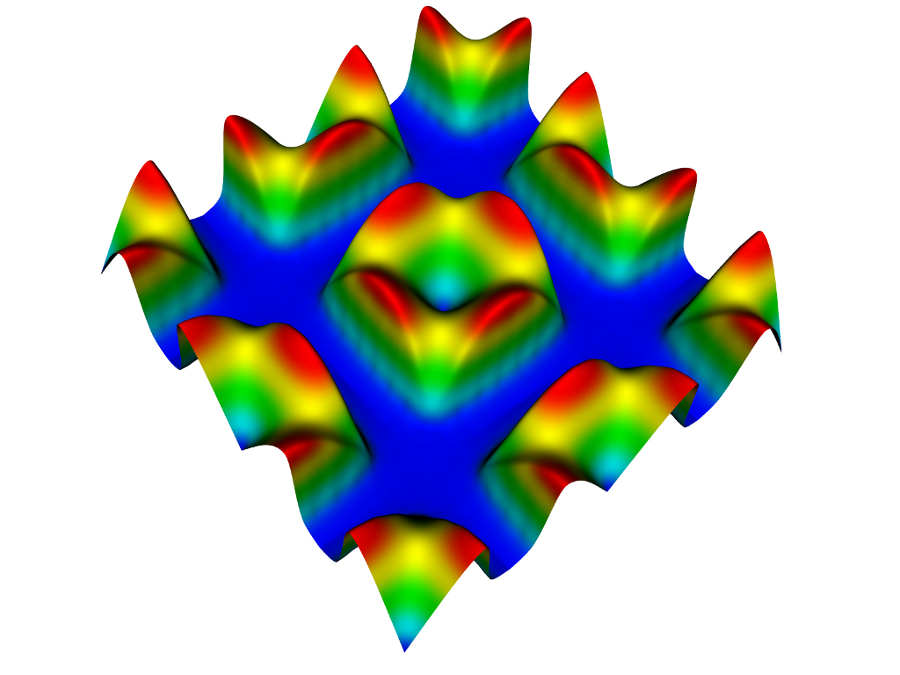}
\includegraphics[width=0.23\linewidth,clip=true]{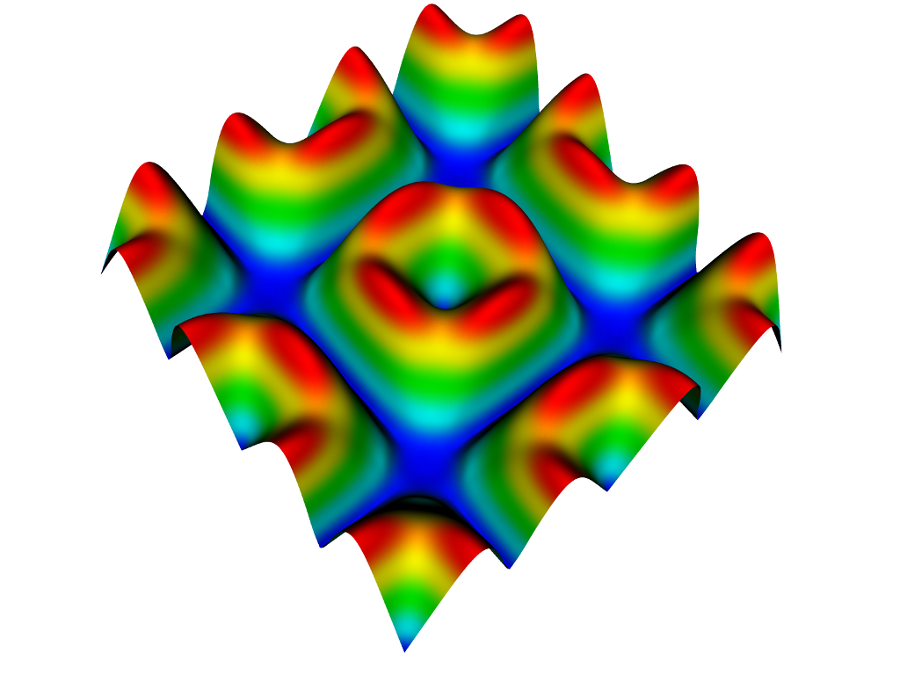}
\includegraphics[width=0.23\linewidth,clip=true]{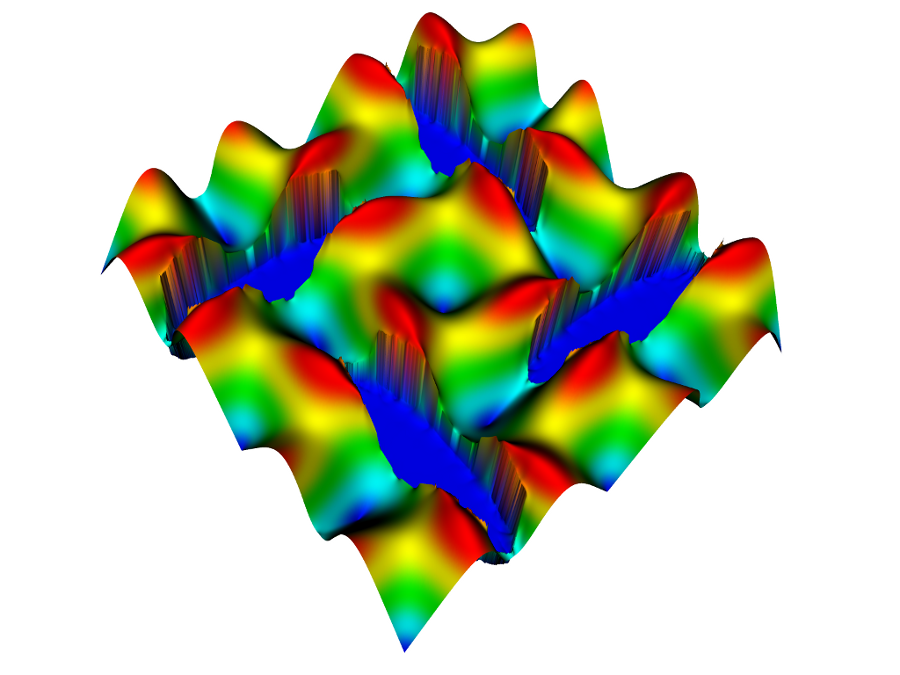}
\includegraphics[width=0.23\linewidth,clip=true]{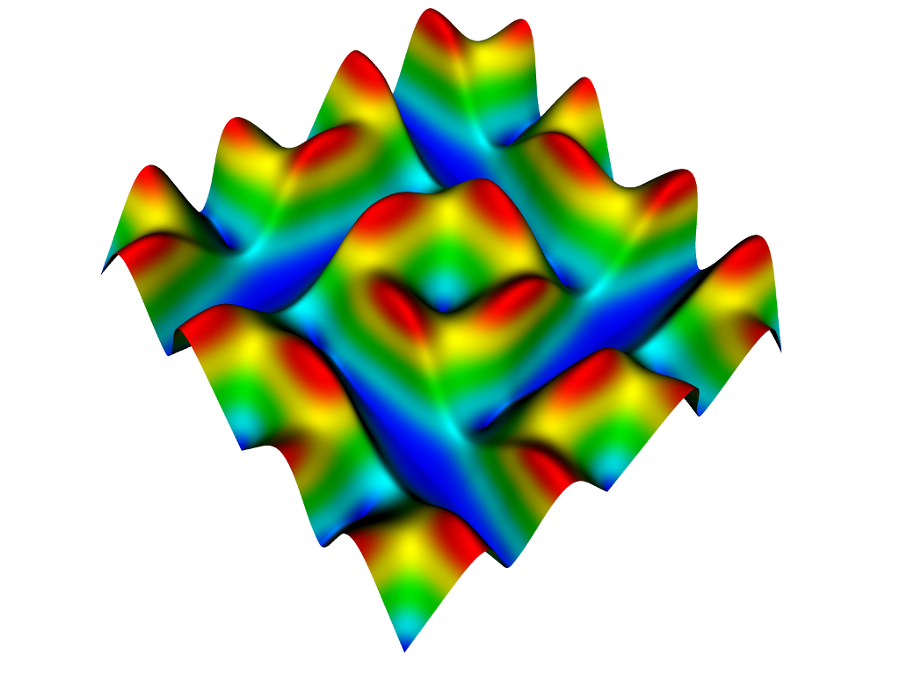}
}
\end{center}
\caption{Electric potential and field values shown in the plane cut from figure \protect \ref{cross_xy}.  From left to right - rectangular, cylindrical, woven and calendered mesh. From top to bottom - electric potential, vertical ($E_z$) and horizontal ($E_{xy}$) electric field values. All cases are for the mesh pitch of 96 $\mu$m and amplification to drift field ratio of 320. Maximal electric field values are shown by red color and in the case of the rectangular mesh component $E_{xy}$ has the lowest value among all 4 geometries. Scaling it so that rectangular maximum value of the $E_{xy}$ component is equal to 1, the other geometries have the following maximal values - cylindrical 1.06, woven 1.09, and calendered mesh 1.19, which agrees with our simple explanation of symmetry caused $E_{xy}$ component cancellation. Even though cylindrical mesh might seem equally symmetrical as the rectangular one the difference is that because of the curvature of the cylindrical wires one gets more of the deflection $E_{xy}$ component directly above the wire which is not the case for the rectangular mesh in which the surface of the wire is not curved - i.e. is flat. Maximum $E_z$ component values are again scaled to the rectangular mesh value of 1.0 - cylindrical 1.18, woven 1.10, and calendered mesh 1.21. }
\label{cross_xy_fld_pot}
\end{figure*}

\subsection{Effect of dielectric spacer}
In practical realization of the micromesh detectors, a set of dielectric spacers is required to keep a constant gap between the mesh and the anode.
The object used as a spacer is typically a dielectric cylinder and in some cases it can have a capillary shape (hollow cylinder). We have studied two types of horizontal spacers made of fused silica which has dielectric constant of $\epsilon=3.5$. In one case the spacer is a full cylinder made of dielectric material and in the other it has a capillary shape of a hollow cylinder, as shown in figure \ref{hollow_spacer}.
The RH code can handle dielectric materials as well as the conductors. After finding the charge densities on the metal parts and polarizations on the dielectric surfaces (RH can handle only homogeneous dielectrics) we can calculate potential and electric field values in any point in space. The cross section of the potential and electric field values along the plane from figure \ref{potential} are given in figure \ref{spacer_impact_potential}. 
Results for the electric field when the spacer is present are quite expected. Introducing a full dielectric cylinder causes larger perturbation in the field resulting in the increased field values in the regions where cylinder touches the electrodes (anode and the mesh). The explanation for this is the following. Introducing the homogeneous dielectric into the region where the electric field is present results in the reduction of the electric field value in the interior of the dielectric object \cite{private_BU}. This can be explained by the polarization charges at the surface of the dielectric. The larger is the volume in which electric field is reduced, the larger is the surface polarization on the dielectric. On the other hand that surface polarization on the dielectric causes increased electric field close to the surface outside the dielectric. This is the reason that the electric field is more increased in the case of the full cylinder than in the case of the capillary spacer. Capillary spacer reduces the electric field produced by electrodes only in its interior which has a much smaller volume than in the case of the full cylinder spacer. Increase in the electric field values is not welcome in the detector design because it can lead to electric discharge (spark). Therefore the dielectric material should occupy the smallest possible volume while retaining its mechanical properties of separating the mesh from the anode. Also the smaller is the dielectric constant, the weaker is the effect of the field enhancement in the regions of dielectric close to the electrodes.

\section{Conclusions}
In this paper, we demonstrate how a complete design and testing of the micro-mesh detector can be done on a personal computer using a novel BEM algorithm, known as the RH Solver. We prove that the RH Solver is a reliable and efficient tool for a complete detector design with 300,000 triangles on a single CPU.
Owning to the small need for computer resources, we are able to accurately model metal meshes and dielectric objects over an area that is significantly larger than the spatial resolution.
Our results show that the electron transparency of the mesh is very sensitive to its detailed geometry, while the gain is almost insensitive to it.


The RH Solver is easily adaptable for a parallel computing environment, or using graphics processors in order to achieve even more complex and faster calculations \cite{paper_with_Joe}.




\end{document}